\documentclass[11pt,a4paper]{article}
\pdfoutput=1 % if your are submitting a pdflatex (i.e. if you have images in pdf, png or jpg format)
\usepackage{jheppub}	% jheppub includes hyperref,color, natbib, amsmath, amssymb, epsfig, graphicx
\usepackage{mathabx}
\usepackage{tikz}
\usetikzlibrary{positioning,arrows}
\usetikzlibrary{decorations.pathmorphing}
\usetikzlibrary{decorations.markings}
\usetikzlibrary{snakes}
\usetikzlibrary{matrix}

\usepackage{epsfig}
\usepackage{graphicx}
\usepackage[vcentermath,enableskew]{youngtab}

\usepackage[utf8]{inputenc}
\usepackage[english]{babel}
\usepackage{makeidx}
\usepackage{tikz}
\tikzset{every picture/.style={line width=0.75pt}} %set default line width to 0.75pt
\tikzset{line/.style={thick, decorate, draw=black,}}
\usepackage{amsfonts}
\usepackage{enumerate}
\usepackage{mathrsfs}
\usepackage{tensor}
\usepackage[autostyle]{csquotes}
\usepackage{subfig}

  \usepackage{multirow}

\newcommand{\ft}[2]{{\textstyle\frac{#1}{#2}}}
\newcommand{\nn}{\nonumber}

\def\be{\begin{equation}}
\def\ee{\end{equation}}
\def\bea{\begin{align}}
\def\eea{\end{align}}
\def\beaq{\begin{eqnarray}}
\def\eeaq{\end{eqnarray}}

 \def\b{\beta}

\title{Tidal resonances for fuzzballs}

\author{Giorgio~Di~Russo
\footnote{Address after February 9th: Institut de Physique Th\'eorique, Universit\'e Paris-Saclay, CEA, Orme des Merisiers, Gif-sur-Yvette, 91191 CEDEX, France }
}
\author{Francesco Fucito,}
\author{Jose~Francisco~Morales}
\emailAdd{giorgio.dirusso@roma2.infn.it}
\emailAdd{fucito@roma2.infn.it}
\emailAdd{morales@roma2.infn.it}

\affiliation{Dipartimento di Fisica, Universit\`a di Roma ``Tor Vergata" \& Sezione INFN Roma2, Via della ricerca scientifica 1, 00133, Roma, Italy}

\abstract{ We  study the gravitational tidal response of D1D5, Top Star and (1,0,n) strata horizonless geometries. We find that the tidal interactions in fuzzball geometries, unlike in the case of black holes, exhibits a sequence of resonant peaks associated to the existence of metastable bound states. The spectrum of resonant frequencies is computed by semi-analytical and numerical methods.}

\begin{document}
\maketitle
\flushbottom
\section{Introduction}
Black holes (BH) are objects which have always been of extreme interests for physicists since their "discovery" as solutions of Einstein's equations at the beginning of the past century. For many years they have been at the center of an intense theoretical research which could have been considered a mathematical amusement until their indirect detection \cite{Webster:1972bsw,Bolton:1972bun}. Nowadays they have become real physical objects \cite{EventHorizonTelescope:2019dse} and there are reasonable hopes that their properties could be measured experimentally via the detection of gravitational waves \cite{LIGOScientific:2016aoc,TheLIGOScientific:2016src,Abbott:2020tfl,Abbott:2020mjq,Abbott:2020khf}.

The main feature of BH's is specified in their name: they are black. They absorb the impinging radiation. Nothing can escape the horizon which cloakes a classical space-time singularity. This is essentially the content of the weak cosmic censorship conjecture \cite{Penrose:1962ij,Penrose:1964wq,Penrose:1969pc,Wald:1997wa}. But even if the geometry outside the horizon shows no pathologies the same cannot be said for the part inside the horizon. The solution to the latter problems might require quantum mechanics which, in turn, raises more problems like the information paradox see for example \cite{Hartle:1996rp} and references therein. These problems lead to a more radical viewpoint questioning the existence of an horizon. Or, in other words, is it possible to distinguish BH's from exotic objects more massive than neutron stars but without an horizon? These objects have been collectively dubbed  Exotic Compact Objects (ECO). This category comprises gravastars, wormholes, firewalls and fuzzballs just to name a few, see \cite{Mathur:2005zp,Cardoso:2016rao,Cardoso:2017cqb} for recent reviews.
ECO's can be distinguished from BH's by their photon sphere shapes \cite{Bianchi:2018kzy,Bianchi:2020des,Bianchi:2020yzr}, multipolar structure \cite{Bena:2020see,Bianchi:2020bxa,Bena:2020uup,Bianchi:2020miz,Mayerson:2020tpn,Bah:2021jno}, quasi normal modes (QNM) and their characteristic echo based gravitational waves (GW) emission \cite{Ikeda:2021uvc}. See \cite{Cardoso:2019rvt}
for a review.

  The internal structure of a celestial object (a BH or an ECO) can be obtained studying their reaction under a gravitational perturbation. The mathematical tool to do so is the tidal Love/dissipation number (TLN) which were introduced in \cite{Love}, see also \cite{Flanagan:2007ix,Damour:2009vw,Binnington:2009bb}. They were shown to vanish for Schwarzschild and Kerr BH's \cite{Damour:2009vw,Binnington:2009bb,Fang:2005qq} but they are non trivial in higher dimensions, non asymptotically flat spaces, in alternative theories of gravity  \cite{Kol:2011vg,Hui:2020xxx,Pereniguez:2021xcj,Cardoso:2017cfl,Emparan:2017qxd,Cardoso:2018ptl} and for BH-like compact objects \cite{Chakraborty:2023zed,Piovano:2022ojl}. The TLN response of a gravity object codified the mixing between the solutions of the wave equation  for a choice of boundary conditions, and it is imprinted in the Post Newtonian (PN) expansion of the gravitational wave signal radiated by a lighter mass inspiraling around it  \cite{Fucito:2023afe}.  It shows up at order 5PN  beyond the quadrupole approximation,  making hard its experimental measure. Aim of this work is to show that, in the case of fuzzballs, the TLN response exhibit resonant peaks in the line of frequencies amplifying the signal in a  significant way.
  The study of TLN resonances will be the central focus of this paper. To carry on this task we will exploit the correspondence recently developed in   \cite{Aminov:2020yma,Bonelli:2021uvf,Bianchi:2021xpr,Bianchi:2021mft,Bonelli:2022ten,Consoli:2022eey} that relates gravity to $N=2$ supersymmetric SU(2) gauge theories and/or two-dimensional conformal field theories (CFT), see also  \cite{Bianchi:2021yqs,Bianchi:2022wku,Bianchi:2022qph,Aminov:2023jve,Bianchi:2023sfs,Fucito:2023afe,Bautista:2023sdf}. This correspondence exploits the results of Seiberg-Witten (SW) \cite{Seiberg:1994rs}, localization \cite{Nekrasov:2002qd,Flume:2002az,Bruzzo:2002xf} and AGT duality \cite{Alday:2009aq}, to provide a combinatorial description of the wave functions describing the gravity systems at linear order.   In this framework, the Post-Newtonian expansion of the gravitational wave emitted by a particle orbiting around a BH is described by a an instanton sum in a quiver gauge theory, or alternatively as a linear combination of conformal blocks of a two-dimensional CFT  \cite{Fucito:2023afe}. The tidal response is given by a ratio of Gamma functions depending on a single
  holomorphic function: {\it the quantum SW period} $\mathfrak{a}(u,q)$  with Coulomb branch parameter $u$ and coupling  $q$ parametrizing the
  orbital number and the wave frequency in gravity \cite{Consoli:2022eey}.

   In this paper, we apply these ideas to the study of the tidal response for D1D5 \cite{Lunin:2001fv}, Topological Star \cite{Bah:2020ogh,Bah:2020pdz} and D1D5p superstrata geometries of type (1,0,s) \cite{Bena:2015bea,Bena:2017xbt}. Perturbations around these geometries are described by a differential equation of the Heun type and can be mapped to a $SU(2)$ gauge theory with matter transforming in the fundamental representation \cite{Bianchi:2021xpr,Bianchi:2021mft,Bianchi:2022qph,Bianchi:2023sfs}. The crucial difference with respect to the case of BH's is the presence of poles in the connection matrix  associated to the real frequencies where the argument of a Gamma function, in the tidal function, becomes a negative integer. We compare the spectrum of resonant frequencies with that of QNM's, i.e.  solutions of the wave equation satisfying regular boundary conditions at the origin and  outgoing ones at infinity.
   QNM's exist for discrete of choices of frequencies  the complex plane. In the case of BH's, where the effective potential has only a maximum (at the photonsphere), they have typically a significantly large imaginary part leading to fast damping modes. Fuzzballs instead allow also for  QNM frequencies with small imaginary part describing  {\it metastable bound states} leaving near the minimum of the effective potential. We find that tidal interactions always blow up at these slowly damping QNM frequencies, providing a gravitational wave picture of the fuzzball interior.

     This is the organization of the paper: in Section 2 we review the gauge-gravity correspondence and illustrate the ideas in a toy model
  of the wave dynamics.  In Section 3, 4 and 5 we discuss the cases of: D1D5 fuzzball, the (1,0,s) Strata and Top Star geometries respectively.
  In each case we study the spectrum of tidal resonances and compare them against that of QNM's.

\section{ TLN vs QNMs}

Black holes and ECO's are often described by integrable geometries where the angular and radial motions of the gravitational perturbations can be separated and described in terms of a ordinary differential equations of Schr\"odinger type
 \be
 \Psi''(z) +Q(z)\Psi(z)=0 \label{eqcan}
 \ee
  with $z$ a radial or angular variable. For example, asymptotically $AdS_4$ Kerr-Newman BH's are described by Heun equations of type (\ref{eqcan}) with four regular singularities located at the BH horizons. Asymptotically flat BH's in four and five dimensions and a large family of D-brane bound states and fuzzballs are instead described by confluent Heun equations obtained by colliding two or more singularities at a point
  (typically zero or infinity).

 The general solution can be always written as
 \be
 \Psi (z)= \sum_{\alpha=\pm}  c_\alpha(\omega) \Psi_\alpha(z)
   \label{psildef}
 \ee
with  $\Psi_\alpha(z)$ given in terms of Heun functions and $c_\alpha$ determined by the boundary conditions at the origin (the horizon for BH's).
We will always set the boundary at $z=1$,  the radial infinity at $z=0$, and choose the solutions  $\Psi_\alpha(z)$ such that
 \be
   \Psi_\pm (z)  \underset{z,{\omega\over z} \to 0}{\approx} z^{{1\over 2} \pm  \left( {1\over 2} +{\ell \over D-3} \right)} (1+\ldots)
   \ee
  with $D$ the dimension and $\ell$ the orbital quantum number of the wave. The limit $\omega/z,z \to 0$ describes the region ({\it near zone}) where distances are large with respect to the size of the object but still much smaller than the gravitational wavelength.  In this limit, centrifugal forces dominate over the rest of the interactions.
  The growing  and decreasing  components, $\Psi_-$  and $\Psi_+$ respectively,   can be viewed accordingly as the ``source ''  and ``response"    terms of the perturbation and the tidal response function ${\cal L}(\omega)$ can be defined as
 \be
  {\cal L}(\omega) ={c_+(\omega)\over c_-(\omega)}
  \ee
     The real and imaginary parts of ${\cal L}(\omega)$ compute the dynamical Love and dissipation numbers of the geometry. The coefficients $c_\pm$ will be computed using the Heun connection formulae derived in \cite{Bonelli:2022ten,Consoli:2022eey}, linking the behaviour of the Heun functions near the origin $z=1$ to the radial infinity $z=0$.

     QNM's on the other hand, are defined as outgoing modes in the opposite asymptotic region where $z\ll \omega$. They
  are in general complex, and exist for discrete choices of the frequencies.

  \subsection{Tidal response resonances vs QNMs}

  We will show that for compact objects ${\cal L}(\omega)$ has an infinite number of poles where the tidal response of the geometry blows up.  This sequence of resonances will be related to the existence of metastable states (QNM's) confined in the interior of the photon sphere. They are characterised by QNM frequencies with small imaginary parts describing bound states living
  near the minimum of the effective potential. QNM frequencies will be computed  by direct integration of the differential equation. They correspond to zeros of the Wronskian,
      $\Sigma_{\rm QNM} (\omega)$,  computed between the solution   $\Psi_{\rm in}(z)$
    satisfying regular boundary conditions at the origin and $\Psi_{\rm out}(z)$ satisfying outgoing boundary conditions at infinity
      \be
   \Sigma_{\rm QNM} (\omega)=\Psi_{\rm in}(z_*) \Psi'_{\rm out}(z_*) -\Psi'_{\rm in}(z_*) \Psi_{\rm out}(z_*)=0 \label{sqnm}
   \ee
     with $z_*$ an arbitrary point. We remind that the Wronskian is constant, so the result  does not depend on the choice of $z_*$. Since both
     the origin and infinity are always singular points, boundary conditions have to be imposed slightly off these points.
     The two solutions are approximated around these points by a series expansion that are used to fix the boundary conditions and are then
       extrapolated numerically to the interior point $z_*$ using Mathematica.

     Alternatively, QNM frequencies can be  estimated relying on a WKB approximation of the solution. In this framework the frequencies $\omega_{n}$   are determined by the quantization condition
\be
\int_{r_1}^{r_2}\sqrt{Q(r,\omega_{n})}dr=\pi\left(n+{1\over2}\right) \qquad , \qquad n=0,1,2,\ldots
\label{wkb}
\ee
with $r_{1,2}$ two zeros of $Q(r,\omega_{n})$ around its maximum. We notice that, in general, the zeros $r_{1,2}$ depend on $\omega_{n}$, so equation (\ref{wkb}) is highly non-trivial, but it can still be solved numerically.

\subsection{Gauge gravity correspondence}

The Heun equation arises also in the study of gauge theories on curved spacetimes. For example,
 the dynamics of  the ${\cal N}=2$ supersymmetric $SU(2)$ gauge theory with four  hypermultiplets living in a Nekrasov-Shatashvili  \cite{Nekrasov:2009rc} $
 \Omega$-background  with parameters $\epsilon_1=1$, $\epsilon_2=0$ is codified in the quantum differential equation
 \be
 \left[    P_L(- z\partial_z+\ft12  ) - P(-z\partial_z  ) z^{-1}+q P_R( -z\partial_z-\ft12 ) z^{-2}\right] W(z)=0 \label{qswcurve}
 \ee
  with $q$ the gauge coupling and
  \beaq
 P_L(x) &=&   (x-m_1)(x-m_2) \qquad , \qquad  P_R(x)= (x-m_3)(x-m_4) \nn\\
 P(x)&=& x^2-u+q\left(  x^2+u +\ft12 -(x+\ft12)\sum_i m_i +\sum_{i<j} m_i m_j \right)
 \eeaq
  Here $u$ parametrises  the Coulomb branch and $m_i$ the masses. This gauge theory can be described by pairs of D4 branes stretched between two NS5 branes or between a NS5 and infinity on the left and on the right. The zeros of $P_{L,R}(x)$, $P(x)$ give the positions of the D4-branes in the three regions and parametrise the four masses and the Coulomb branch. We will label the hypermultiplet content by $N_f=(2,2)$ and its decoupling limits (where some masses are sent to infinity) by $N_f=(N_L,N_R)$.
%We introduce the superfluous parameter $\delta=0,1$ for later convenience, but we stress that only the product $P_L(x)P_R(x)$ is physically relevant.
     Equation (\ref{qswcurve}) can be always brought to the form (\ref{eqcan}) by taking
  \be
     W(z) = z^{1-{m_3+m_4\over 2}} (1-z)^{-{m_1+m_2+1\over 2}}  (z-q)^{{m_3+m_4-1\over 2}} \Psi(z)
   \ee
   leading to
   \beaq
   Q_{22} (z) &=& \frac{\frac{1}{4}-
   \left(\ft{m_1+m_2}{2}\right){}^2}{(z-1)^2}+\frac{\frac{1}{4}-
   \left(\ft{m_3-m_4}{2}\right){}^2}{z^2}+\frac{\frac{1}{4}- \left(\ft{m_3+m_4}{2}\right){}^2}{(z-q)^2}\nn\\
  &&\qquad   +\frac{2 m_1 m_2+m_3^2+m_4^2-1}{2 (z-1)
   z}+\frac{(1-q)U
 }{(z-1) z (z-q)} \label{Q22}
   \eeaq
   with
   \be
   U= u+\ft14 -\ft{1}{2} \left(m_3^2+m_4^2\right)
   -\frac{q \left(1-m_1-m_2\right) \left(1-m_3-m_4\right)}{2
   (1-q)}
   \ee
   The gauge theory can be alternatively described by a mirrored version of (\ref{qswcurve}) given by
   \be
 \left[  q  P_L(- \tilde{z}\partial_{\tilde{z}}  +\ft12  ) - P(-\tilde{z}\partial_{\tilde{z}}  ) \tilde{z}^{-1}+ P_R( -\tilde{z}\partial_{\tilde{z}}-\ft12 ) \tilde{z}^{-2}\right] \widetilde W( \tilde{z})=0 \label{qswcurve2}
 \ee
   after the identifications
   \be
   z=q \tilde z \qquad , \qquad \widetilde{Q}_{22}(\tilde{z})=q^2 \,Q_{22}(q \tilde{z} ) \qquad , \qquad \widetilde{W}(\tilde{z})=W( q \tilde{z} )
   \ee
    The use of this second picture will become clear later.
   Using any of the two equivalent descriptions, the dynamics of the gauge theory is codified
   into a single holomorphic function  $\mathfrak{a}(u,q)$, known as the Seiberg Witten quantum period. To compute it, it is convenient to Fourier transform (\ref{qswcurve}), or  (\ref{qswcurve2}),  and bring it into the form of a difference equation \cite{Poghossian:2010pn,Fucito:2011pn}. The integrability condition of this difference equation can be written in the infinite fraction form \cite{Poghosyan:2020zzg}
  \beaq
\label{fractionequality}
\Sigma_{\rm TLN}(\mathfrak{a})\equiv \frac{ q M(\mathfrak{a}+1)}{P(\mathfrak{a}+1)-\frac{ q M(\mathfrak{a}+2)}{P(\mathfrak{a}+2)-\ldots
}}
+\frac{ q M(\mathfrak{a})}{P(\mathfrak{a}-1)-\frac{ q M(a-1)}{P(\mathfrak{a}-2)-\ldots
}}-P(\mathfrak{a})=0
\eeaq
with
\be
M(x)=\prod_{i=1}^4 (x-m_i-\ft12)
\ee
(\ref{fractionequality}) can be easily solved for $\mathfrak{a}(u,q)$ order by order in $q$.
 The starting point of the recursion is the free theory $q= 0$, where  $\mathfrak{a} \underset{q\to 0}{\approx} \sqrt{u}$.
 In this limit, the wave equations reduce to
\beaq
\Psi (z) &  \underset{q\to 0}{\approx} & \sum_{\alpha=\pm} c_\alpha  z^{\frac{1}{2}{+}\alpha \mathfrak{a} } (1{-}z)^{\frac{1}{2} \left(1{-}m_1{-}m_2\right)} \,
   _2F_1\left(\ft{1}{2}{+}\alpha\mathfrak{a}{-}m_1,\ft{1}{2}{+}\alpha \mathfrak{a}{-}m_2 ; 1{+}2\alpha \mathfrak{a};z\right)  \label{f210}\\
   \widetilde{\Psi} (z) &  \underset{q\to 0}{\approx} &   \widetilde{c}_3  z^{\frac{1}{2}{+}{m_4-m_3\over 2} } (1{-}z)^{\frac{1}{2} \left(1{-}m_3{-}m_4\right)} \,
   _2F_1\left(\ft{1}{2}{-}m_3{+}\widetilde{\mathfrak{a}},\ft{1}{2}{-} m_3{-}\widetilde{\mathfrak{a}} ; 1{+}m_4{-}m_3;z\right)
 {+}(3\leftrightarrow 4)  \nn
\eeaq
where $c_\pm, \tilde c_{3,4}$ are some constants.  They are determined by requiring regularity at the origin $z=1$.

Expanding the hypergeometric functions around $z=1$, assuming that the masses are real\footnote{We will later see that this is always the case for the fuzzball geometries considered here.} and $m_1+m_2>0$,
$m_3+m_4>0$, one finds that  regularity at the origin requires
\be
\Psi   \underset{z\to 1}{\sim}   (1-z)^{1+m_1+m_2\over 2} \qquad {\rm or} \qquad  \widetilde{\Psi}   \underset{z\to 1}{\sim}  (1-z)^{1+m_3+m_4\over 2}
\ee
 This fixes the ratios between the two asymptotic coefficients around $z=0$ to be
 \beaq
{ \cal L} &=&  {c_+\over c_-}= \frac{\Gamma \left(-2    \mathfrak{a} \right)
\Gamma \left(\frac{1}{2}+ m_1 + \mathfrak{a}
	\right) \Gamma \left(\frac{1}{2}+  m_2 +\mathfrak{a} \right)
	}{\Gamma \left(2   \mathfrak{a} \right)  \Gamma \left(\frac{1}{2}+  m_1 - \mathfrak{a}
	\right) \Gamma \left(\frac{1}{2}+ m_2 - \mathfrak{a} \right)}
	 \nn\\
\widetilde{ \cal L} &=&  {\tilde c_3\over \tilde c_4}= \frac{\Gamma \left(1+m_3-m_4\right)
\Gamma \left(\frac{1}{2}+ m_4 + \mathfrak{a}
	\right) \Gamma \left(\frac{1}{2}+  m_4 -\mathfrak{a} \right)
	}{\Gamma \left(1-m_3+m_4 \right)  \Gamma \left(\frac{1}{2}+  m_3+ \mathfrak{a}
	\right) \Gamma \left(\frac{1}{2}+ m_3 - \mathfrak{a} \right)}
	\label{theta}
 \eeaq
 Turning on $q$, the differential equation becomes a Heun equation and the $c$'s coefficients become the components of the  connection matrix relating two basis of Heun functions. The crucial observation in \cite{Consoli:2022eey} is that the ratio between the c's coefficients in the Heun case is given again by (\ref{theta}) with $\mathfrak{a} \underset{q\to 0}{\approx} \sqrt{u}$ replaced by the quantum SW period  $\mathfrak{a}(u,q)$. The whole non-triviality of the Heun connection matrix is codified into a single holomorphic function
 $\mathfrak{a}(u,q)$.

 The confluence limits of the Heun equation can be studied similarly by decoupling some  masses in the gauge theory description.
 Notice that this operation breaks the Left-Right symmetry, so which description you use makes a difference. We find it convenient to use
  ${ \cal L}(\omega)$ or $\widetilde{ \cal L}(\omega)$ to describe the tidal response for theories with $N_L>N_R$ or $N_L<N_R$.
For example, the $N_f=(2,1)$ is obtained in the limit $q\to 0$, $m_{4} \to \infty$, keeping finite their product $q m_4 \to -q$. In this limit (\ref{qswcurve})
  keeps its form with the replacement $P_L(x) \to (x-m_3)$.  The differential equation becomes a confluent Heun equation. A further decoupling of the $m_3$ mass produces a   $N_f=(2,0)$ theory, described by a reduced confluent Heun equation.  Similarly one can decouple the $m_{1,2}$ masses, using the formulae with the tilde leading to $N_f=(1,2)$ and $N_f=(0,2)$ gauge theories.

   We will be mainly interested in the poles of the tidal functions. They appear for frequencies such that the argument  of some of the gamma functions in (\ref{theta}) becomes a negative integer. In the case of BH's this never happens, since the masses are always purely imaginary and $\mathfrak{a}$ is real.
    We find instead that the tidal response of fuzzball geometries exhibits typically an infinite sequence of resonances corresponding to
    solutions of
  \beaq
 \ft12+m_{1,2}+\mathfrak{a}    &=&-n \qquad , \qquad n=0,1,2,\ldots  \nn\\
\ft12+m_{4}\pm \mathfrak{a}    &=&-n \qquad , \qquad n=0,1,2,\ldots  \label{an}
 \eeaq
 We find that typically only one of the two equations admits real solutions. To find the frequencies, we will plug $\mathfrak{a}$ obtained from (\ref{an}) into (\ref{fractionequality}) and look numerically  for the zeros of $\Sigma_{\rm TLN}(\omega)$ along the $\omega$-line.

\subsection{Toy wave dynamics }

\begin{figure}[htbp]
\centering
\includegraphics[width=0.45\textwidth]{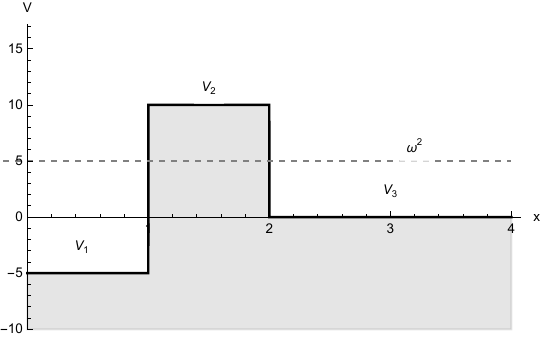}\includegraphics[width=0.45\textwidth]{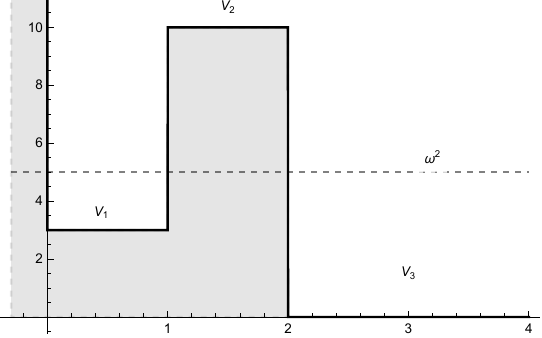}
\caption{ Effective potential $V(x)$ for the BH (left) and fuzzball (right) cases}\label{graphVeff}
\end{figure}

 In this Section we illustrate the ideas of tidal resonances and their connection with QNM's in a toy model of the wave dynamics introduced in \cite{Bianchi:2023sfs}.   We consider the quantum mechanics of a particle moving in a piecewise constant potential $V(x)$ with different values
 in the intervals  $(0,L ]$, $   \left[ L,L+\Delta \right]$ and $\left[ L+\Delta,\infty\right]$,
  see figure \ref{graphVeff}.   The  quantum probability $\Psi(x)$ of finding the particle at position $x$ is determined by
 the Schr\"odinger equation (\ref{eqcan})
with
\be
 Q(x,\omega)=\omega^2-V(x) \label{qomega}
 \ee
and $\omega^2$ the particle energy.
The general solution of (\ref{eqcan}) can be written as
 \be\label{wavfunc2}
\Psi(x) =\left\{\begin{array}{lll}
       {A }\cos k_1 x + \tilde{A} \sin k_1 x  &~~~~~~~~&   0< x\leq L\\
        B e^{-\beta x}+   C e^{\beta x}  &&   L< x\leq L+\Delta\\
         D e^{i k_3 x} +\tilde{D} e^{-i k_3 x}  &&   x>L+\Delta
        \end{array} \right.
 \ee
 with
 \be
   k_1=\sqrt{\omega^2-V_1} \quad , \quad \beta=\sqrt{V_2-\omega^2} \quad , \quad k_3=\omega
 \ee
 We consider the two extreme cases where the boundary of the space at $x=0$ represents either a BH (ingoing wave only) or a fuzzball (perfectly reflecting mirror).
 They correspond to the choice of boundary conditions
 \beaq
 \Psi_{\rm Fuzz}(x) & \underset{x\to 0}{\approx}&  \sin{ k_1 x}   \nn\\
 \Psi_{\rm BH}(x) & \underset{x\to 0}{\approx} &  e^{-{\rm i} k_1 x}
 \eeaq
  We can view the B (decreasing) and C (growing)  components in the intermediate region in (\ref{wavfunc2}) as describing the amplitude of
  the ``response'' of the geometry to a ``source" perturbation.
  The TLN is defined as the ratio of the response and source coefficients
  \be
{\cal L}(\omega)={B(\omega)\over C(\omega)}
\ee
 Matching functions and derivatives at $x=L$ one finds
   \beaq
{\cal L}_{\rm Fuzz} (\omega)&=&  {\b \tan k_1 L - k_1 \over \b \tan k_1 L + k_1} e^{2 \b L}\nn\\
{\cal L}_{\rm BH} (\omega)&=&   {\b + i k_1 \over \b - i k_1} e^{2 \b L } \label{ll11}
\eeaq
 The important difference we observe already at this level, is that the BH TLN has no poles since $\beta={\rm i} k_1$ has no real solutions; while  ${\cal L}_{\rm Fuzz} (\omega)$ exhibits an infinite number of poles located at the solutions of the trascendental equation:
\be
 \tan k_1 L =-{k_1\over \beta} \label{metalove}
\ee
QNM's instead are defined as solutions involving only outgoing waves at infinity, i.e. $\tilde{D}=0$. Matching  functions and derivatives at $x=L,L+\Delta$ one finds that QNM frequencies are determined by the eigenvalue equations
\beaq \label{qnmeig}
{\rm Fuzz}: \qquad  && \frac{\beta}{k_1}\tan{k_1 L}=\frac{i k_3\tanh{\beta\Delta}-\beta }{\beta\tanh{\beta \Delta}-i k_3 }  \nn\\
{\rm BH}: \qquad  &&\frac{ i \beta}{k_1}=\frac{i k_3\tanh\beta\Delta-\beta}{\beta\tanh\beta \Delta-ik_3 } \label{ll22}
\eeaq
 We notice that the equations defining the poles of  ${\cal L}_{\rm Fuzz} (\omega)$ in (\ref{ll11}) are very different in general from
those in (\ref{ll22}) defining QNM's.  Still in the limit $\beta \Delta \to \infty$,  the two conditions coincide. Indeed in this limit the right hand side of the first equation in (\ref{ll22}) reduces to $-1$, leading again to (\ref{metalove}). The   resonances of the tidal function are therefore associated to  the QNM's describing the metastable states confined inside the cavity.

\section{D1-D5 circular fuzzball}

\begin{figure}[t]
  \begin{minipage}{\textwidth}
    \centering
    \includegraphics[width=0.48\textwidth]{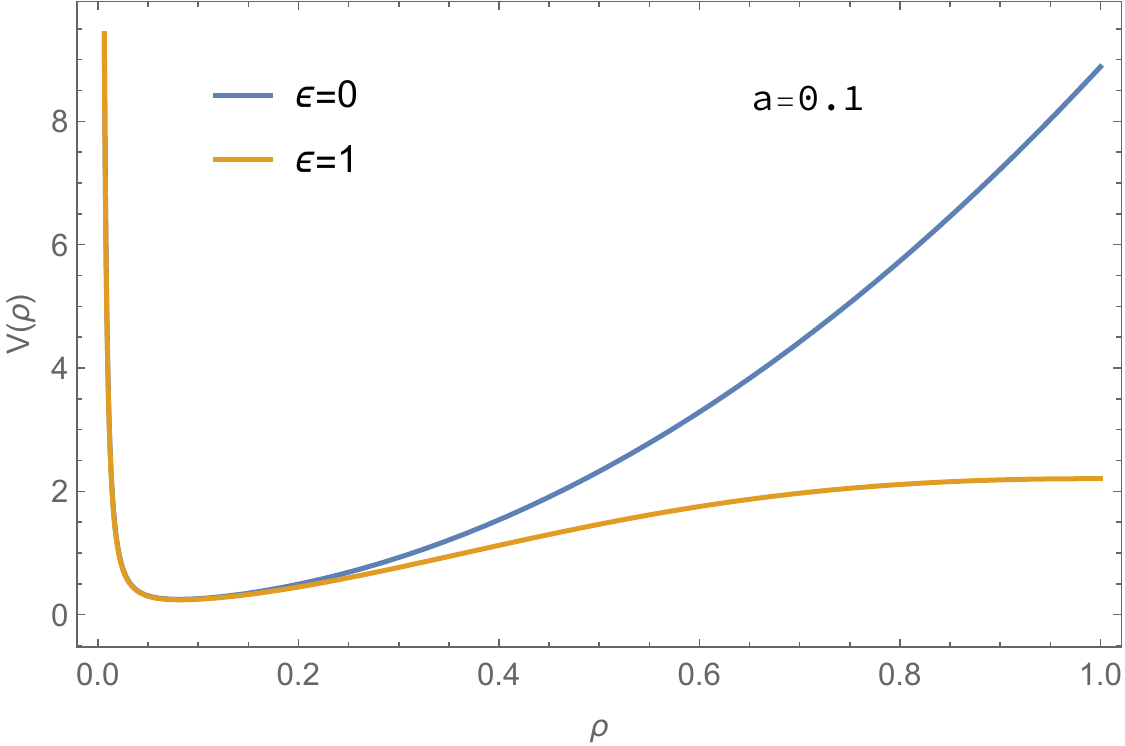} \includegraphics[width=0.48\textwidth]{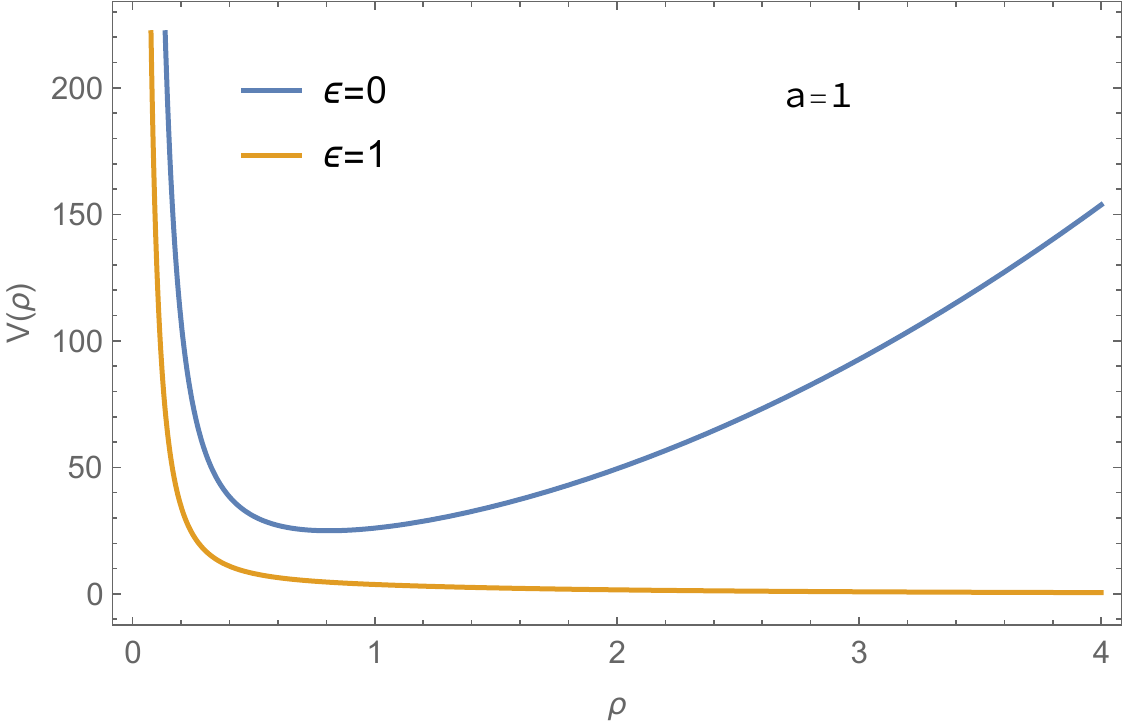}
    \caption{ D1D5 Effective potential: $\ell=m_\psi=2$, $m_\phi=\mu=0$,  $L=1$.}
    \label{figv}
  \end{minipage}
\end{figure}

The D1D5 circular fuzzball is a smooth solution of minimal six-dimensional supergravity specified by a circular profile of radius $a$.
The six-dimensional metric is given by
\be
ds^2 = H^{-1}  \left[ -2 (du+\beta)(dv+\gamma) +  H^2 ds_4 \right]
\ee
with
\beaq
ds_4 &=& (\rho^2+ a^2 \cos^2\theta)\left( {d\rho^2\over \rho^2+a^2}  +d\theta^2\right) +(\rho^2+a^2)  \sin^2\theta\, d\phi^2+  \rho^2 \cos^2\theta\, d\psi^2  \nn\\
\beta &=& { a  L^2(\sin^2\theta d\phi{-}\cos^2\theta d\psi)  \over  \sqrt{2}  \Sigma } ~ , ~ \gamma = { a L^2  (\sin^2\theta d\phi{+}\cos^2\theta d\psi)  \over  \sqrt{2}  \Sigma }  \nn\\
  H  &=& \epsilon{+}  {L^2 \over  \Sigma} \quad, \quad
\Sigma = \rho^2+ a^2 \cos^2\theta \quad, \quad  t={u+v\over \sqrt{2} }  \quad , \quad y={u-v\over \sqrt{2}}  %\qquad, \qquad  R ={L^2 \over \sqrt{2} a }
\eeaq
 We have introduced the auxiliary parameter $\epsilon=0,1$ that interpolates  between asymptotically $AdS_3\times S^3$ and flat geometries respectively. We consider a massive scalar perturbation satisfying
 \be
 (\Box-\mu^2) \Phi =0
 \ee
The scalar wave equation is separable using the ansatz
\be
\Phi=R(r) S(\chi) e^{{\rm i} ( - \omega  t + m_\psi \psi + m_\phi \phi  )} \label{phiw}
\ee
The angular and radial equations read
\begin{align}
 \partial_\chi \left[ \chi (1-\chi^2) S'(\chi )\right] & +  \chi S(\chi )\left(-\frac{m_\phi^2}{1-\chi^2}-\frac{m_\psi^2}{\chi ^2}{-}\epsilon a^2 (1-\chi^2) ( \omega^2  \epsilon  {-}  \mu^2   ) {+}A  \right)=0 \label{d1d5wave}\\
 \partial_\rho \left[  \rho (\rho^2{+}a^2)  R'(\rho )\right]& {+}  \rho R(\rho )  \left[  \frac{ \left(am_\phi{-} L^2 \omega  \right)^2}{\rho^2{+}a^2}{-}\frac{a^2  m_\psi^2}{\rho ^2} {+}\rho^2\epsilon(\epsilon\omega^2{-}\mu^2) {-}(L^2{+}a^2\epsilon)\mu^2{+}\right.\\\nn
&\left.+\epsilon\omega^2(2L^2{+}a^2\epsilon){-}A)   \right] =0 \nn
\end{align}
with $\chi=\cos\theta$.
The two equations can be written in the normal  form (\ref{eqcan}) with
\begin{align}\label{effv}
Q_\chi(\chi)=&a^2 \epsilon  \left(\mu ^2-\omega ^2 \epsilon \right)+\frac{2
   A-2 m_{\psi }^2-m_{\phi }^2+3}{4 (\chi +1)}+\frac{-2 A+2
   m_{\psi }^2+m_{\phi }^2-3}{4 (\chi -1)}+\frac{1-4 m_{\psi
   }^2}{4 \chi ^2}+\\\nn
   &+\frac{1-m_{\phi }^2}{4 (\chi
   -1)^2}+\frac{1-m_{\phi }^2}{4 (\chi +1)^2}\\\nn
   Q_\rho(\rho)=&\epsilon(\epsilon \omega^2-\mu^2)+{{1\over4}-m_\psi^2\over \rho^2}-{1+A-m_\psi^2+L^2(\mu^2-2\epsilon \omega^2)\over \rho^2+a^2}+{(L^2\omega-a m_\phi)^2-a^2\over (\rho^2+a^2)^2}
\end{align}
In figure \ref{figv} we display the effective potential $V(\rho)$ governing the radial motion, for some values of the parameters.
We restrict ourselves to the case $m_\phi=0$, where $Q_\rho(\rho) \sim \omega^2 -V(\rho)$. We observe that for $a$ small enough the potential
for, both
the asymptotically flat and the near horizon geometry, has always a minimum so it allows for the existence of metastable bound states.  This is not the case for asymptotically flat D1D5 fuzzball with $a$ large.
 To build the gauge gravity dictionary, it is convenient to move the singularities to the points $0$, $1$ and $\infty$ and write $Q$ in the form
 \be\label{swform20}
Q_{20} (z) =-\frac{q}{z^3}+ \frac{\ft{1}{4}- \left(\ft{m_1-m_2}{2}\right){}^2}{(z-1)
   z}+\frac{\ft{1}{4}- \left(\ft{m_1+m_2}{2}\right){}^2}{(z-1)^2
   z}+\frac{u-\frac{1}{4}}{(z-1) z^2}
 \ee
    For the angular equation one finds the gauge gravity dictionary
\beaq\label{angdicfuzz}
z_\chi& =& {1\over \chi^2}  \qquad,\qquad  S(\chi)={ \Psi(z) \over \sqrt{(1-z) }} \\
q_\chi &=& { \epsilon a^2\over 4} ( \epsilon {\omega}^2 {-} \mu^2)  \quad , \quad
m_{1,2}^\chi =  {m_\phi\pm m_\psi\over2}\quad,\quad  u_\chi ={1{+}A\over 4} \nn
\eeaq
 while radial variables are given by
\begin{align}
z &={ a^2\over \rho^2+ a^2}  \quad, \quad   R(r)={ \Psi(z) \over \sqrt{(1-z) }} \quad,\quad  u ={1\over 4}[1{+}A{+}L^2(\mu^2{-}2\epsilon\omega^2){+}a^2\epsilon(\mu^2{-}\epsilon\omega^2)] \label{dicrgen}\\
 q& ={ \epsilon a^2\over 4} ( \mu^2{-}\epsilon {\omega}^2)   \qquad, \qquad
m_1 ={m_\psi+m_\phi\over 2} - {L^2\omega\over 2a} \qquad,\qquad m_2 ={m_\psi-m_\phi\over 2} +{L^2\omega \over 2a}  \nn
\end{align}
 The characteristic polynomials are
  \beaq
 P_L(x) &=&   (x-m_1)(x-m_2)    \qquad , \qquad  P_R(x)=1   \nn\\
  P(x) &=&  x^2-u+q   \qquad , \qquad M(x)=  (x-m_1-\ft12) (x-m_2-\ft12) \label{p02}
 \eeaq
 In the following we consider in turn the cases  of asymptotically $AdS_3\times S^3$ $\epsilon=0$
     and asymptotically flat $\epsilon=1$ geometries.
    The crucial difference between the two cases is that the wave equation for $\epsilon=0$ is of hypergeometric type while that
    for $\epsilon=1$ is given by a reduced confluent Heun equation with an irregular singularity at $z=0$.

 \subsection{ Near horizon geometry }
 \label{sectiond1d5e0}

 We start by considering the simple case $\epsilon=0$. For this choice, the gauge theory representing the fuzzball perturbations is free $q=0$ and
  both angular and radial equations take a hypergeometric form with regular singularities at $0,1,\infty$. Their solutions can therefore be written in terms of hypergeometric functions.
    In particular, the angular function $S(\chi)$ can be written in terms
  of  5d spherical harmonics with separation constant
  \be
  A=\ell(\ell+2)
  \ee
 and
  \be
  \mathfrak{a}=\sqrt{u}= \ft12 \sqrt{ (\ell+1)^2 {+}  \mu^2L^2}
  \ee
      The tidal response function ${\cal L}(\omega)$ is given by (\ref{theta}) with resonant frequencies  located at the solutions of (\ref{an}) .
    Using the gauge gravity dictionary  (\ref{dicrgen}), one finds that the solutions with $\omega$  real exist only for those special values of $m_1$ leading to
  \be
  \omega^{\rm poles}_n={a \over L^2} (2 n+2 \mathfrak{a}+1+m_\psi+m_\phi) \quad, \quad n=0,1,2,\ldots \label{we0}
  \ee
 Beside the sequence of poles, the tidal response has an infinite number of zeros located at
  \be
  \omega^{\rm zeros}_n={a \over L^2} (2 n-2 \mathfrak{a}+1+m_\psi+m_\phi) \quad, \quad n=0,1,2,\ldots \label{we1}
  \ee
Finally we observe that in the massless limit $\mu \to 0$, the sequences of zeros and poles accidentally fall on top of each other leading to a finite
result for ${\cal L}$. Indeed setting $\mu=0$ from the very beginning and expanding for small $z$ one finds the logarithmic divergent tidal response
    \be
    R(z) \underset{z\to \infty} {\sim} z^{ -\ell-{1\over 2}} \left[  (1+\ldots)  +  z^{2\ell+2}  \log z\, {\cal L}_{\rm reg} \, (1+\ldots )  \right]
    \ee
     with
     \beaq
      {\cal L}_{\rm reg}= {\rm Res}_{\mathfrak{a}=\ft{\ell+1}{2}}   {\cal L}(\mathfrak{a})
   &=&  -\frac{    \prod_{i=0}^{\ell}  \left[  m_1^2- (i+\ft12)^2 \right] \left[  m_2^2- (i+\ft12)^2 \right] }{(2 \ell+2)!  (2 \ell+1)!}
     \eeaq
  a polynomial function in $\omega$ with no resonances. This can be observed in the left plot of figure \ref{D1D5e0} where no resonances show up
  for $\epsilon=0$.

    \subsection{Asymptotically flat geometry}

    \begin{figure}[t]
\centering
\includegraphics[width=0.45\textwidth]{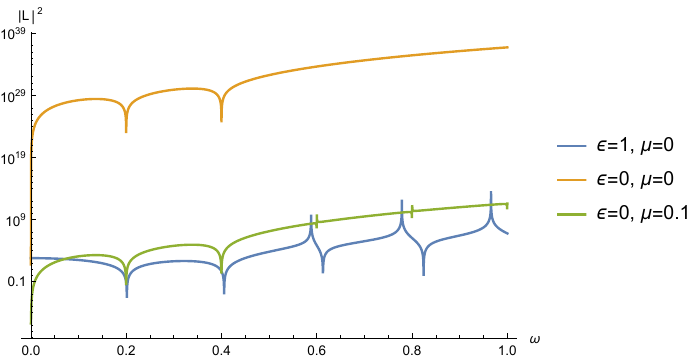}\includegraphics[width=0.45\textwidth]{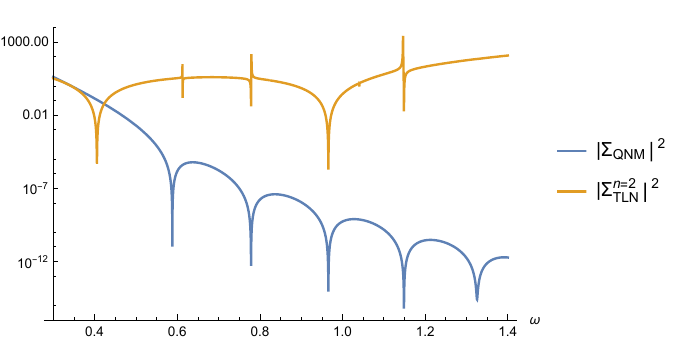}
\caption{D1D5 fuzzball $\ell=m_\psi=2$, $m_\phi=0$, $a=0.1$, $L=1$, $p=0$}\label{D1D5e0}
\end{figure}
   \begin{table}[t]
\centering
\begin{tabular}{|l|l|l|l||l|l|}
\hline
n & $\omega^{\rm TLN }_{\rm poles}$ &$ \omega^{\rm QNM}_{\rm WKB}$  & $  \omega^{\rm QNM}_{\rm num}$ &A & $\omega^{\rm TLN }_{\rm poles,extra}$   \\
\hline
 0 & 0.5881 &0.586058  & $0.5881  $ &  8.00086 & 0\\
 1 &0.778579  & 0.774457 & $0.778579  $ & 8.00152  & 0.201356 \\
 2 &0.965754  & 0.960428  & $0.965751   $ &  8.00233  & 0.405563 \\
 3 & 1.14872 &1.14282  & $1.14873  $ &  8.00330 & 0.612965 \\
 4 & 1.32444 &1.31985  & $1.32553  $ & 8.00438  &  0.824199\\
  \hline
\end{tabular}
\caption{ TLN resonance vs QNM frequencies for D1D5 fuzzballs with $L=\epsilon=1$, $\ell=m_\psi=2$,  $m_\phi=\mu=0$, $ a=0.1$. }
\label{tablel2}
\end{table}

    \begin{table}[h]
\centering
\begin{tabular}{|l|l|l|l||l|l|}
\hline
n & $\omega^{\rm TLN }_{\rm poles}$ &$ \omega^{\rm QNM}_{\rm WKB}$  & $  \omega^{\rm QNM}_{\rm num}$ &A & $\omega^{\rm TLN }_{\rm poles,extra}$   \\
\hline
 0 &2.15643  &2.1563  &2.15643  & 120.004 & 0\\
 1 &2.34803  &2.34771 & 2.34803  & 120.005& 0.200365  \\
 2 &2.53883  & 2.53835 &2.53883   &  120.005& 0.401467 \\
 3 &2.72882  &  2.7282 &2.72882  & 120.006& 0.603316\\
 4 & 2.91796& 2.91721 & 2.91796 &120.007  & 0.805929 \\
  \hline
\end{tabular}
\caption{ TLN resonance vs QNM frequencies for D1D5 fuzzballs with $L=\epsilon=1$, $\ell=m_\psi=10$,  $m_\phi=\mu=0$, $ a=0.1$. }
\label{tablel10}
\end{table}

  For  $\epsilon=1$ the singularity at $z=0$ is irregular and the solutions
  $\Psi_\alpha$ are given in terms of the reduced confluent Heun functions rather than hypergeometric ones. The tidal response is described by
    ${\cal L}(\omega)$ in  (\ref{theta}), with  $\mathfrak{a}(q)$ determined by the continuous fraction equation (\ref{fractionequality}). We refer the reader to \cite{Bianchi:2022qph} for recent investigation of the Love number of circular D1D5 fuzzballs.
  To determine the resonant frequencies we first plug
 \be
\mathfrak{a}=-n- \ft12-m_{1}  \qquad , \qquad n=1,2,\ldots \label{an3}
  \ee
  into (\ref{fractionequality}) truncated to a given instanton number $q^k$. We also set $\mathfrak{a}^\chi=1+\ft{\ell}{2}$
  in the continuous fraction equation associated to the angular equation.
   We then use the dictionaries (\ref{angdicfuzz}) and (\ref{dicrgen}) to express the angular and radial equations as a system of two algebraic equations and solve it numerically for $A$ and $\omega$.
    For example, truncating to  zero instanton order, one finds $\mathfrak{a}\approx \sqrt{u}+\ldots $ that after using the gauge gravity dictionary
     (\ref{dicrgen})  leads to $A\approx \ell(\ell+2)+\ldots $ and
  \be
   \omega_n \approx {a\over L^2} \left( 2n+\ell+m_\psi+m_\phi +2\right)+\ldots
  \ee
  We notice that at this order the result coincides with that of the near horizon geometry given by  (\ref{we0}).
   This is not  surprising, since for $a$ small the minimum of the effective potential is located deep inside the near horizon geometry which
   provides a good description of the physics of metastable states.
  Higher instanton corrections can be incorporated by keeping more and more terms in the continuous fraction equations.
We find two solutions for the $m_1$-dependent pole equation for each $n$  and no one for the equation involving $m_2$.
The results converge very fast, reaching four-five digits accuracy already at three or four instanton order for $a=0.1$.
  The results for the frequencies $\omega$  for some specific choices of parameters are displayed in table \ref{tablel2}-\ref{tablel10}.  The results are compared against those based on the WKB method and direct numerical integration of the differential equation\footnote{QNM frequencies are characterised also by a very small negative imaginary part  that will be omitted here. }.  The match between QNM frequencies and poles of the tidal response is impressive. The agreement can be also appreciated in the right plot
  of figure \ref{D1D5e0} where we display the characteristic functions $|\Sigma_{\rm QNM}|^2$ and $|\Sigma_{\rm TLN}|^2$  defined by (\ref{sqnm}) and (\ref{fractionequality}) respectively.  QNM frequencies and resonant peaks of the tidal response correspond to the zeros of these two functions and they perfectly match.

   In the opposite limit, let us say $a=1$ (in units of L), the effective potential has no minima, so metastable states are not present.  For such choices, the resonances of the tidal interactions still exist but they are no longer related to  QNM's.  It would be interesting to understand what kind of information about the internal structure of the fuzzball can be extracted from the spectrum of resonances in this case.

 \section{Superstrata (1,0,s) }

Superstrata \cite{Bena:2015bea,Bena:2017xbt} generalize the D1D5 circular geometries by introducing a third charge and allowing a more general string profile. The class we consider here, labelled by (1,0,s), is characterized by two real positive numbers $a$, $b$ and an integer $s$.
We focus on the near horizon region, where the  geometry is integrable  \cite{Bena:2017upb},  and therefore radial and angular motion can be separated.   The geometry is asymptoticaly $AdS_3\times S^3$ and it is described by the metric
\be
ds^2 = -{2 \over \sqrt{ \cal P}}  (dv+\beta)\left[ du+w + {{\cal F}_s\over 2} (dv+\beta)\right]   + \sqrt{ \cal P}  ds_4  \label{metric10n}
\ee
with $ t=\ft{u+v}{\sqrt{2}}$, $y=\ft{u-v}{\sqrt{2}}$ and
\beaq
\beta &=& { a^2 R_y  \over  \sqrt{2}  \Sigma }  (\sin^2\theta d\varphi-\cos^2\theta d\psi)  \qquad, \qquad {\cal F}_s=-{b^2\over a^2} \left( 1-{\rho^{2s}\over (\rho^{2}+a^{2})^s} \right) \nn\\
w &=&  { a^2 R_y  \over    \sqrt{2}   \Sigma } (\sin^2\theta d\varphi+\cos^2\theta d\psi -{\cal F}_s \sin^2\theta d\varphi )    \nn\\
 {\cal P} &=&  {L^4\over \Sigma^2} \left(1- { a^2 b^2 \rho^{2s} \sin^2\theta \over (2a^2+b^2) (\rho^2+a^2)^{s+1} } \right)  \quad , \quad
  R_y ={L^2 \over \sqrt{a^2+\ft{b^2}{2} } } \label{ryab}
 \eeaq
 The solution is specified by three independent parameters chosen between $R_y$, $L$, $a$ and $b$ satisfying the relation (\ref{ryab}).
 Starting  from the ansatz
 \be
\Phi=R(r) S(\chi) e^{{\rm i} ( -\omega t+ m_\varphi \varphi + m_\psi \psi  )} \label{phiw2}
\ee
 the massive wave equation separates into \cite{Bena:2017upb}
 \beaq
&&\partial_\chi  \left[ \chi \left(1-\chi ^2\right)
   S'(\chi ) \right]   +\chi \left( A   -\frac{m_\phi^2}{1-\chi ^2}-\frac{m_\psi^2}{\chi ^2} \right) S(\chi) =0 \nn\\
   && \partial_\rho \left[ \rho \left(a^2+\rho ^2\right) R'(\rho) \right]  +\rho \left[  -(A{+}\mu^2 R_y^2) {+} \frac{\left(  2 a^2 m_\varphi- 2 a^2 R_y \omega - b^2 \, R_y  \, \omega\right)^2 }{ 4(\rho^2+a^2) } \right.  \label{eqr10n}\\
&&\left. -\frac{a^2  m_\psi^2}{\rho ^2}  +
\frac{b^2  R_y   \omega \rho^{2s} \left(   4 a^2 m_\varphi- 2 a^2 R_y  \omega +b^2  R_y
   \omega \right)}{ 4 a^2(\rho^2+a^2)^{s+1} }  \right] R(\rho)=0 \nn
  \eeaq
 The angular equation  is solved by spherical harmonics in five dimensions with
 \be
 A=\ell(\ell+2)
 \ee
  The radial equation leads to a Fuchsian differential equation with two regular singularities and one irregular one of degree $1+[s/2]$. We consider
  here the cases of s=0,1,2 leading to  hypergeometric or Heun equations.

  The tidal response of superstrata geometries are holographically related to two-point functions in the dual two-dimensional CFT living on the boundary of $AdS_3$ at the intersection of D1D5 branes.
  Resonances of the tidal response therefore correspond to poles of the two point function and have been recently studied in \cite{Giusto:2023awo}.

      \begin{figure}[t]
\centering
\includegraphics[width=0.5\textwidth]{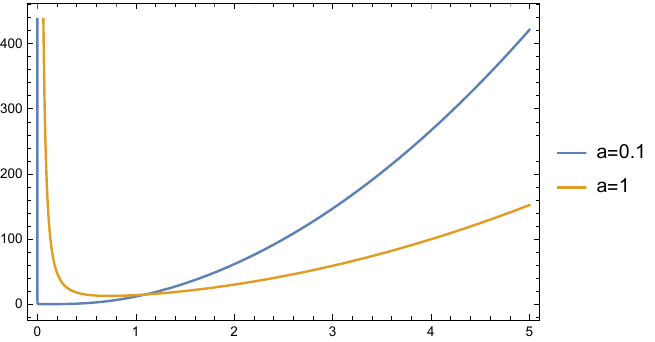}
\includegraphics[width=0.33\textwidth]{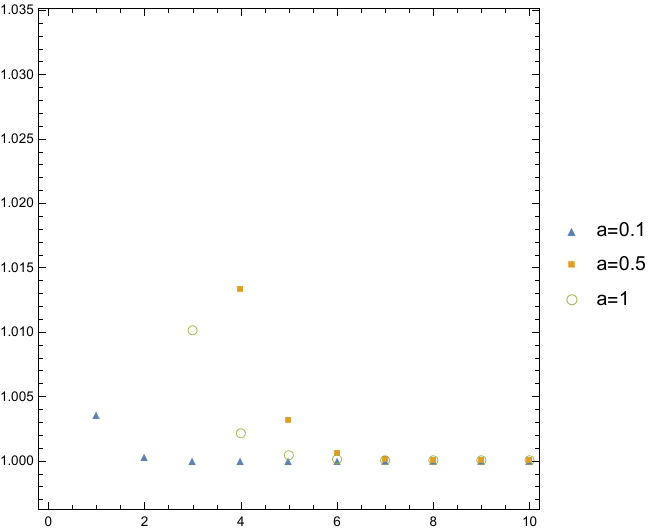}
\caption{(1,0,1) superstratum: convergence for fixed $\ell=m_\psi=2$, $m_\phi=0$, $b=R_y=1$  for various $a$ Left) Effective potential. Right)
Convergence of the instanton series. }\label{}
\end{figure}

 \subsection{$(1,0,0)$ stratum}

  The $(1,0,0)$ geometry describes a D1D5 two charge state with a string profile  partially  wrapping an internal circle. Its size and orientation is specified by the parameters $a$ and $b$.  The case $b=0$ corresponds to the D1D5 circular profile studied in the last section.
  The important difference with respect to the circular case is that for $b\neq 0$, the wave dynamics is separable only for $\epsilon=0$,  so we restrict ourselves to this choice. The (1,0,0) near horizon geometry is described again by a free $N_f=(2,0)$ gauge theory

  The analysis of the tidal response follows mutatis mutandis that of the circular D1D5 case given in Subsection \ref{sectiond1d5e0} and all formulae in that section still hold but now gauge and gravity variables are related to each other by the dictionary
 \beaq
z&=& {a^2\over \rho^2+a^2}  \quad , \quad  R(z)={\Psi(z) \over \sqrt{1-z}} \quad, \quad q = 0 \quad , \quad u=\ft14(\ell+1)^2+\ft{\mu^2 R_y^2}{4}   \\
m_1&=&  \ft{m_\psi}{2}-\ft{1}{2} \sqrt{ (m_\varphi-
     \omega  R_y)^2 +\ft{b^2 \omega ^2 R_y^2}{2a^2}}  \quad , \quad m_2= \ft{m_\psi}{2}+\ft{1}{2} \sqrt{ (m_\varphi-
     \omega  R_y)^2 +\ft{b^2 \omega ^2 R_y^2}{2a^2}} \nn
  \eeaq
   Resonances are found by solving
  \be
 \ft12+m_{1}+\sqrt{u}    =-n \qquad , \qquad n=0,1,2
 \ee
  in $\omega$, leading to
   \be\label{100qnm}
\omega^{\rm TLN}_n=\frac{2 a^2}{\left(2 a^2+b^2\right) R_y} \left[m_{\varphi }+ \sqrt{\left(\ft{b^2}{2 a^2}+1\right)
   \left(2 n+m_{\psi }+\tilde \ell+2\right)^2-\ft{b^2 m_{\varphi }^2}{2
   a^2}}\right]
     \ee
   with
   \be
   \tilde{\ell}=\sqrt{(\ell+1)^2 +\mu^2 R_y^2} -1 \label{tildel}
     \ee
     Like in the near horizon D1D5 case, no resonances show up when $\mu=0$ due to an accidental cancellation between the numerator and denominator of the response function.

\subsection{(1,0,1) microstates}

    \begin{figure}[t]
\centering
A)\includegraphics[width=0.45\textwidth]{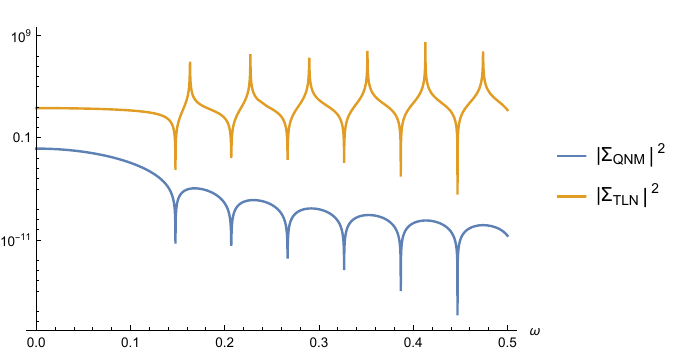}\includegraphics[width=0.45\textwidth]{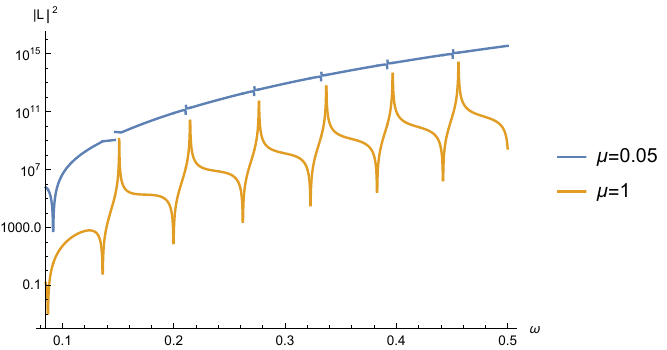} \\
B) \includegraphics[width=0.45\textwidth]{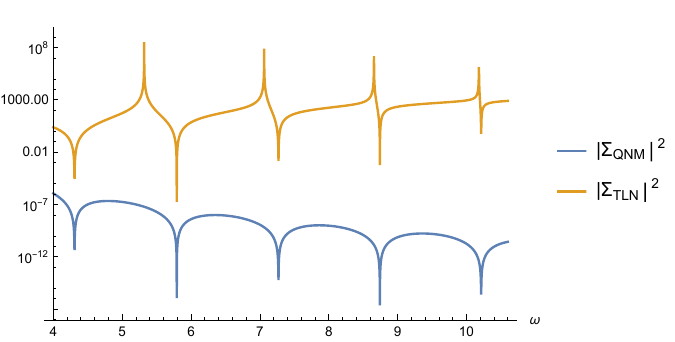}\includegraphics[width=0.45\textwidth]{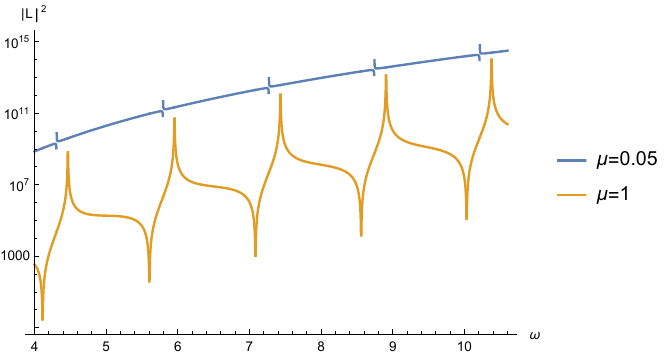}
\caption{(1,0,1)-Tidal response: $\ell=m_\psi=2$, $R_y=b=1$, A) $a=0.1$, B) $a=1$. }\label{fig101}
\end{figure}

Microstate geometries of type (1,0,1) are described by an interacting $SU(2)$ gauge theory with $ N_f=(0,2)$ fundamentals.
  The wave equation can be put into the normal form (\ref{eqcan}) with
\be
\widetilde{Q}_{02}=-\frac{q}{z}+\frac{u+\ft{1}{4}- \ft{1}{2} \left(m_3^2+m_4^2\right)}{(1-z)
   z}+\frac{\ft{1}{4}-
   \left( \ft{m_3-m_4}{2}\right){}^2}{z^2}+\frac{\ft{1}{4}-
   \left( \ft{m_3+m_4}{2}\right){}^2}{(z-1)^2}
\ee
  and
\beaq
z&=& {a^2\over \rho^2+a^2}  \quad , \quad  R(z)={\Psi(z) \over \sqrt{1-z}} \quad ,\quad
   u=  \left({ m_{\varphi } \over 2} -{ \omega  R_y\over 4a^2}(2a^2+b^2) \right)^2  \\
q&=&  \frac{b^2
   \omega  R_y}{16 a^4} \left(\omega  \left(2 a^2+b^2\right) R_y-4 a^2 m_{\varphi
   }\right) \quad, \quad
m_3 = \ft{1}{2} \left(m_{\psi }-\tilde{\ell}-1\right) \quad , \quad m_4= \ft{1}{2}
   \left(m_{\psi }+\tilde{\ell}+1\right) \nn
  \eeaq
  with $\tilde{\ell}$ given by (\ref{tildel}).
  The characteristic polynomials are
  \beaq
 P_L(x) &=& 1   \qquad , \qquad  P_R(x)= (x-m_3)(x-m_4)      \nn\\
  P(x) &=&  x^2-u+q   \qquad , \qquad M(x)=  (x-m_3-\ft12) (x-m_4-\ft12) \label{p02}
 \eeaq
  The tidal response is described by $\widetilde{\cal L}(q)$ given in (\ref{theta}).
   As in the case of the near horizon D1D5 circular geometry  a small mass is introduced in order to  split the sequences of poles and zeros of $\widetilde{\cal L}$.   We use $\mu$ as a regulator, assuming it is always small but not zero.
   The wave dynamics is described again by a reduced confluent Heun equation, but now the irregular singularity is located at $z=\infty$.
   On the other hand the point $z=0$ describing the space-time infinity, corresponds now to a regular singularity.

  Assuming $m_\psi=m_3+m_4>0$ and noticing that $m_3<m_4$, one finds the resonances at the frequencies  satisfying the quantization conditions
  \be
 \ft12+m_{4}-\mathfrak{a}=-n  \qquad , \qquad n=1,2,\ldots \label{an2}
  \ee
 At zero instanton this leads to
\be
\omega_n  \approx \frac{2a^2}{R_y(2a^2+b^2)} \left( 2n+m_{\psi }+m_\varphi+\widetilde{\ell}+2 \right)   + \ldots
\ee
Higher instanton contributions can be incorporated as well.
 The convergence rate of the instanton series is shown in figure \ref{fig101}. A $0.1 \%$ of accuracy is reached already with four instantons for $a=0.1$ or six for $a=1$.  In tables \ref{tablea1} we display the results for the resonant frequencies and their associate QNM's for some choices of the gravity parameters. Again a perfect agreement is found between resonances of the tidal response and QNM frequencies.

 \begin{table}[t]
\centering
\begin{tabular}{|c|c|c|c|}
\hline
n & $\omega^{\rm TLN }_{\rm poles}$ &$ \omega^{\rm QNM}_{\rm WKB}$  & $  \omega^{\rm QNM}_{\rm num}$   \\  \hline
0 & 0.14776 & 0.14747 & 0.14776 \\ \hline
1 & 0.207013 & 0.206229 & 0.207013 \\ \hline
2 & 0.266735 & 0.265636 & 0.266735 \\ \hline
3 & 0.326705 & 0.325385 & 0.326705 \\ \hline
4 & 0.386816 & 0.385332 & 0.386816 \\ \hline
\end{tabular}
\quad
\begin{tabular}{|c|c|c|c|}
\hline
n & $\omega^{\rm TLN }_{\rm poles}$ &$ \omega^{\rm QNM}_{\rm WKB}$  & $  \omega^{\rm QNM}_{\rm num}$   \\  \hline
0 & 4.308 & 4.29375 & 4.308 \\ \hline
1 & 5.79275 & 5.76227 & 5.79275 \\ \hline
2 & 7.26969 & 7.2296 & 7.26969 \\ \hline
3 & 8.74252 & 8.6961 & 8.74252 \\ \hline
4 & 10.2129 & 10.162 & 10.2129 \\ \hline
\end{tabular}
\caption{(1,0,1) TLN resonances for $\ell=m_\psi=2$, $m_\phi=\mu=0$, $R_y=b=1$: Left) $a=0.1$ and Right)  $a=1$}
\label{tablea1}
\end{table}

\subsection{(1,0,2) microstates}

\begin{figure}[t]
\centering
\includegraphics[width=0.5\textwidth]{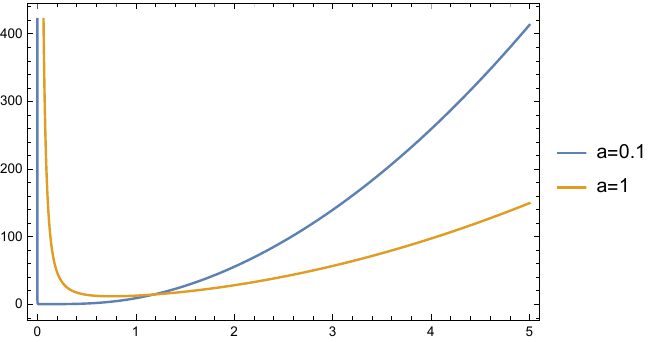}
\includegraphics[width=0.33\textwidth]{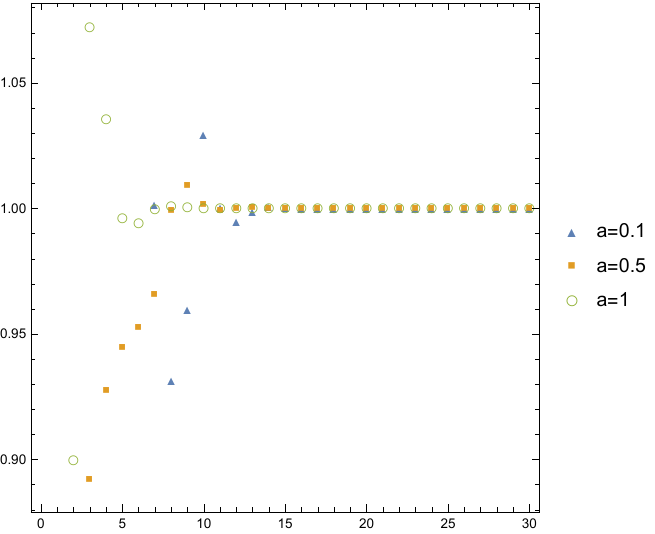}
\caption{(1,0,2) superstratum: convergence for fixed $\ell=m_\psi=2$, $m_\phi=0$, $b=R_y=1$. Left) Effective potential.  Right) Convergence of the   instanton series}\label{fig102pot}
\end{figure}

   \begin{figure}[t]
\centering
\includegraphics[width=0.45\textwidth]{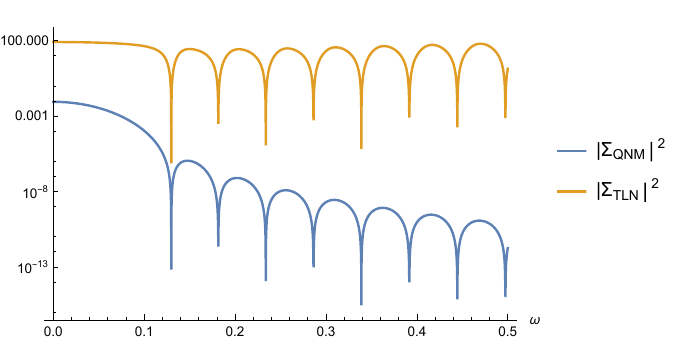}  \includegraphics[width=0.45\textwidth]{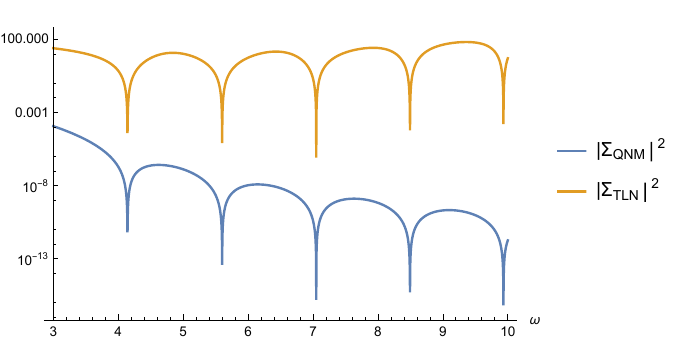}
\caption{(1,0,2)-Tidal response: $\ell=m_\psi=2$, $R_y=b=1$, L) $a=0.1$, R) $a=1$. }\label{fig102}
\end{figure}

 Perturbations of the (1,0,2) geometry are described by a confluent Heun equation related to a $N_f=(1,2)$ gauge theory. The wave equation
can be written in the form (\ref{eqcan}) with
\beaq
\tilde{Q}_{12}&=&\frac{m_2 q}{z}+\frac{\frac{1}{4}-\frac{1}{4}
   \left(m_3-m_4\right){}^2}{z^2}+\frac{\frac{1}{4}-\frac{1}{4}
   \left(m_3+m_4\right){}^2}{(z-1)^2}\nn\\
 &&  \qquad +\frac{\frac{1}{2} \left(m_3 q+m_4 q-m_3^2-m_4^2-q\right)+u+\frac{1}{4}}{(1-z)
   z}-\frac{q^2}{4}   \label{q12}
\eeaq
and gauge gravity dictionary
\beaq
z&=& {a^2\over\rho^2+ a^2} \quad, \quad  R(z)={\Psi(z) \over \sqrt{z(1-z)}} \quad ,\quad
 u=  \left({b^2R_y\omega-2a^2(m_\phi-R_y\omega)\over 4a^2}\right)^2 +{(1-m_\psi)\kappa\over 2}\nn  \\
q &=&  \gamma   \quad , \quad   m_1=  \ft{1}{2} \left(\ell+m_{\psi }+1\right) \quad , \quad m_2= \ft{1}{2}
   \left(-\ell+m_{\psi }-1\right) \quad , \quad m_3= \frac{\kappa }{4}
     \eeaq
   where
 \be
 \kappa  =-{b\sqrt{R_y\omega}\over 2a^2} \sqrt{a^2(4m_\phi-2R_y \omega)-b^2R_y\omega}
 \ee
 The characteristic polynomials are
  \beaq
 P_L(x) &=&  (x-m_2) \qquad , \qquad  P_R(x)= (x-m_3)(x-m_4) \nn\\
 P(x)&=& x^2-u+q\left(   x+\ft12 - m_2-m_3-m_4\right) \qquad , \qquad M=\prod_{i=2}^4 (x-m_i-\ft12)
 \eeaq

 \begin{table}[t]
\centering
\begin{tabular}{|c|c|c|c|}
\hline
n & $\omega^{\rm TLN }_{\rm poles}$ &$ \omega^{\rm QNM}_{\rm WKB}$  & $  \omega^{\rm QNM}_{\rm num}$   \\  \hline
0 & 0.129812 & 0.129697 & 0.129812 \\ \hline
1 & 0.181538 & 0.180922 & 0.181538 \\ \hline
2 &0.233759  &0.232842  & 0.233759 \\ \hline
3 & 0.286251 &0.285131  & 0.286251 \\ \hline
4 & 0.338906 & 0.337637 &0.338906  \\ \hline
\end{tabular}
\quad
\begin{tabular}{|c|c|c|c|}
\hline
n & $\omega^{\rm TLN }_{\rm poles}$ &$ \omega^{\rm QNM}_{\rm WKB}$  & $  \omega^{\rm QNM}_{\rm num}$   \\  \hline
0 &4.14296  & 4.13302 &4.14296 \\ \hline
1 &5.60077  &  5.57343& 5.60077 \\ \hline
2 &7.04986  &7.01223  & 7.04986 \\ \hline
3 &8.49333  & 8.44897 & 8.49333 \\ \hline
4 &9.93312  & 9.88406 &9.93312  \\ \hline
\end{tabular}
\caption{(1,0,2) TLN resonances for $\ell=m_\psi=2$, $m_\phi=\mu=0$, $R_y=b=1$: Left) $a=0.1$. Right)  $a=1$}
\label{tablea1}
\end{table}

 The effective potentials for various choices of $a$ is displayed in figure \ref{fig102pot}.
  Real resonances in the tidal response are obtained at the zeros of the equation in the second line of (\ref{an}) with the choice of a minus sign.
 At zero instanton, one finds
  \begin{align}
\omega_n = \frac{2a^2}{R_y(2a^2+b^2)} \left[m_\phi+\sqrt{(\ell+m_\psi+2n+2)^2-{2b(m_\psi-1)\over2a^3+b^2}(b(m_\psi-1)-\Delta)}\right]
\end{align}
with
\beaq
\Delta^2 &=& {-}2 a^2 \left(m_{\psi }{-}m_{\phi}{+}2 n{+}\ell {+}2\right) \left(m_{\psi }{+}m_{\phi }{+}2 n{+}\ell
   {+}2\right){-}b^2 \left[4 n \left(m_{\psi }{+}n\right){+}2 \ell
   \left(m_{\psi }{+}2 n{+}2\right)+\right. \nn\\
   &&\left.{+}6 m_{\psi }{-}m_{\phi }^2{+}8 n{+}\ell
   ^2{+}3\right]
\eeaq
    Including higher instanton contributions one finds the results listed in table  \ref{tablea1}. Again a perfect agreement is observed between
 tidal response resonances and QNM's. The match is also illustrated in figure \ref{fig102}.

%\FloatBarrier
%\appendix
%
%\section{Hypergeometric Connection formulae}
%\begin{align}
%{}_2F_1(a,b;c;z^{-1})=&e^{-i \pi a}z^a{\Gamma(c)\Gamma(b-a)\over \Gamma(b)\Gamma(c-a)}{}_2F_1(a,a-c+1;a-b+1;z)+\\\nn
%&+e^{-i \pi b}z^b{\Gamma(c)\Gamma(a-b)\over \Gamma(a)\Gamma(c-b)}{}_2F_1(b,b-c+1;-a+b+1;z)\\\nn
%{}_2F_1(a,b;c;z^{-1})=&z^a{\Gamma(c)\Gamma(c-a-b)\over \Gamma(c-a)\Gamma(c-b)}{}_2F_1(a,a-c+1;a+b-c+1;1-z)+\\\nn
%&e^{i \pi (c-a-b)}z^a(1-z)^{c-a-b}{\Gamma(c)\Gamma(a+b-c)\over \Gamma(a)\Gamma(b)}{}_2F_1(1-b,c-b;c-a-b+1;1-z)\\\nn
%{}_2F_1(a,b;c;z)=&{\Gamma(c)\Gamma(c-a-b)\over \Gamma(c-a)\Gamma(c-b)}{}_2F_1(a,b;a+b-c+1;1-z)+\\\nn
%&+(1-z)^{c-a-b}{\Gamma(c)\Gamma(a+b-c)\over \Gamma(a)\Gamma(b)}{}_2F_1(c-a,c-b;c-a-b+1;1-z)
%\end{align}
%
%
\section{Topological stars}

\begin{figure}[t]
\centering
\includegraphics[width=0.5\textwidth]{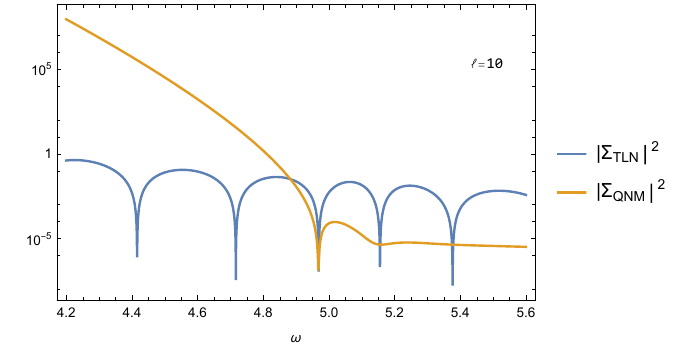}\includegraphics[width=0.5\textwidth]{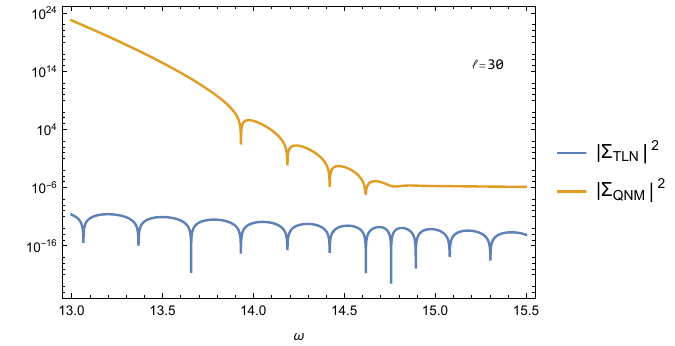}
\caption{Topological star :  $r_b=1$, $r_s=0.8$, $p=0.25$, $\mu=0$ Left) $\ell=10$, Right) $\ell=30$}\label{figts}
\end{figure}

Topological star (TS) are horizonless and asymptotically flat solutions of the Einstein-Maxwell theory in 5 dimensions. The solution is spherically symmetric so the angular equation can be explicitly solved in terms of spherical harmonics.
 The radial equation takes the form
 \be
 {d\over dr}  \left[ \Delta(r) R'(r) \right] +\left[ { \omega^2 r^3 \over r-r_s} -{ p^2 r^3 \over r-r_b}  -\ell(\ell+1) \right] R(r)=0 \label{radialtop}
 \ee
  with $p$ the internal momentum and
\be
\Delta(r)=(r-r_s)(r-r_b) \qquad , \qquad r_s < r_b <2 r_s
\ee
The geometry is smooth and ends on $r_b$.   The massless scalar wave equation on this geometry can be put into the form (\ref{eqcan}) with
 \beaq\label{swform12}
 Q_{21}(z) &=& -{q^2\over 4}{-}{m_3 q\over z}{+}{1{-}(m_1{-}m_2)^2\over 4z^2(1+z)}{+}{1{-}(m_1{+}m_2)^2\over 4z(1+z)^2}{+}{1{-}4u{+}2q(1{-}m_1{-}m_2) \over 4z(1+z)}
\eeaq
 The resulting equation is of confluent like type and describes the dynamics of a supersymmetric SU(2) gauge theory with $N_f=(2,1)$ fundamentals and characteristic polynomials
  \beaq
 P_L(x) &=&  (x-m_1)(x-m_2) \qquad , \qquad  P_R(x)= (x-m_3) \nn\\
 P(x)&=& x^2-u+q\left(  x+\ft12 -\sum_{i=1}^3 m_i \right) \quad , \quad M(x)=\prod_{i=1}^3 (x-m_i-\ft12)
 \eeaq
 Gauge and gravity variables are related by
 \begin{align}\label{raddict}
 z &= { r-r_b \over r_b-r_s } \qquad , \qquad R(r(z) )={ \Psi(z) \over \sqrt{ z(z+1) }}\nn\\
m_{1}& =m_2=-{  \omega r_s^{3\over 2} \over \sqrt{r_b-r_s} }   \quad , \quad  q = -2 {\rm i} (r_b-r_s)\omega  \quad , \quad
 m_3  =-{{\rm i} \omega \over 2} (r_b+2r_s)   \nn \\
u & =\left(\ell{+}\ft{1}{2}\right)^2 {-}{\rm i}(r_b{-}r_s)\omega {-}\omega^2r_s^2 \left(3 {+}2 {\rm i}  \sqrt{ \ft{r_b}{r_s} {-}1}\right)
\end{align}
   The tidal response is given by ${\cal L}$ in (\ref{theta}) and the poles are located at frequencies such that
    \beaq
 \ft12+m_{1}+\mathfrak{a}    &=&-n \qquad , \qquad n=0,1,2,\ldots
 \eeaq
   The results for some choices of the parameters are shown in table \ref{tabts} and figure \ref{figts}. We observe again that the QNM's with a small imaginary part always correspond to resonances in the tidal response of the gravity object. The opposite is not true, since some of the resonances of the tidal response have no counterpart in the spectrum of QNM's. It would be interesting to understand the physical origin of these resonances.

\begin{table}[t]
\centering
\begin{tabular}{|c|c|c|c|}
\hline
n & $\omega^{\rm TLN }_{\rm poles}$ &$ \omega^{\rm QNM}_{\rm WKB}$ & $ \omega^{\rm QNM}_{\rm num}$ \\ \hline
0 &13.9319 &13.8491 & 13.9319 \\ \hline
1 &14.1875 &14.1106 & 14.1875 \\ \hline
2 &14.4203 &14.3514 & 14.4203 \\ \hline
3 &14.619 & 14.5637 & 14.6179 \\ \hline
\end{tabular}
\caption{Topological star $r_b=1$, $r_s=0.8$, $\ell=30$, $p=0$, $\mu=0.25$}
\label{tabts}
\end{table}

\section*{Acknowledgements}
We thank A. Argenzio, Y. F. Bautista, I. Bena, M. Bianchi, G. Bonelli, V. Collazuol, C. Di Benedetto, S. De Angelis, A. Grillo, C. Iossa, N. Kovensky, P. Pani, A. Ruiperez, R. Russo, G. Sudano and A. Tanzini for  fruitful scientific exchanges. G.D.R. thanks IPhT Paris-Saclay for the kind hospitality during completion of this work.
Finally we thank the MIUR PRIN contract 2020KR4KN2 ``String Theory as a bridge between Gauge Theories and Quantum Gravity'' and the INFN project ST\&FI ``String Theory and Fundamental Interactions'' for partial support.

\providecommand{\href}[2]{#2}\begingroup\raggedright\endgroup


\begin{thebibliography}{10}

\bibitem{Webster:1972bsw}
B.~L. Webster and P.~Murdin, \emph{{Cygnus X-1-a Spectroscopic Binary with a
  Heavy Companion ?}}, \href{http://dx.doi.org/10.1038/235037a0}{\emph{Nature}
  {\bf 235} (1972) 37--38}.

\bibitem{Bolton:1972bun}
C.~T. Bolton, \emph{{Dimensions of the Binary System HDE 226868 = Cygnus X-1}},
  \href{http://dx.doi.org/10.1038/physci240124a0}{\emph{Nature Phys. Sci.} {\bf
  240} (1972) 124--127}.

\bibitem{EventHorizonTelescope:2019dse}
{\scshape Event Horizon Telescope} collaboration, K.~Akiyama et~al.,
  \emph{{First M87 Event Horizon Telescope Results. I. The Shadow of the
  Supermassive Black Hole}},
  \href{http://dx.doi.org/10.3847/2041-8213/ab0ec7}{\emph{Astrophys. J. Lett.}
  {\bf 875} (2019) L1}, [\href{http://arxiv.org/abs/1906.11238}{{\tt
  1906.11238}}].

\bibitem{LIGOScientific:2016aoc}
{\scshape LIGO Scientific, Virgo} collaboration, B.~P. Abbott et~al.,
  \emph{{Observation of Gravitational Waves from a Binary Black Hole Merger}},
  \href{http://dx.doi.org/10.1103/PhysRevLett.116.061102}{\emph{Phys. Rev.
  Lett.} {\bf 116} (2016) 061102}, [\href{http://arxiv.org/abs/1602.03837}{{\tt
  1602.03837}}].

\bibitem{TheLIGOScientific:2016src}
{\scshape LIGO Scientific, Virgo} collaboration, B.~P. Abbott et~al.,
  \emph{{Tests of general relativity with GW150914}},
  \href{http://dx.doi.org/10.1103/PhysRevLett.116.221101,
  10.1103/PhysRevLett.121.129902}{\emph{Phys. Rev. Lett.} {\bf 116} (2016)
  221101}, [\href{http://arxiv.org/abs/1602.03841}{{\tt 1602.03841}}].

\bibitem{Abbott:2020tfl}
{\scshape LIGO Scientific, Virgo} collaboration, R.~Abbott et~al.,
  \emph{{GW190521: A Binary Black Hole Merger with a Total Mass of $150 ~
  M_{\odot}$}},
  \href{http://dx.doi.org/10.1103/PhysRevLett.125.101102}{\emph{Phys. Rev.
  Lett.} {\bf 125} (2020) 101102}, [\href{http://arxiv.org/abs/2009.01075}{{\tt
  2009.01075}}].

\bibitem{Abbott:2020mjq}
{\scshape LIGO Scientific, Virgo} collaboration, R.~Abbott et~al.,
  \emph{{Properties and astrophysical implications of the 150 Msun binary black
  hole merger GW190521}},
  \href{http://dx.doi.org/10.3847/2041-8213/aba493}{\emph{Astrophys. J. Lett.}
  {\bf 900} (2020) L13}, [\href{http://arxiv.org/abs/2009.01190}{{\tt
  2009.01190}}].

\bibitem{Abbott:2020khf}
{\scshape LIGO Scientific, Virgo} collaboration, R.~Abbott et~al.,
  \emph{{GW190814: Gravitational Waves from the Coalescence of a 23 Solar Mass
  Black Hole with a 2.6 Solar Mass Compact Object}},
  \href{http://dx.doi.org/10.3847/2041-8213/ab960f}{\emph{Astrophys. J.} {\bf
  896} (2020) L44}, [\href{http://arxiv.org/abs/2006.12611}{{\tt 2006.12611}}].

\bibitem{Penrose:1962ij}
R.~Penrose, \emph{{Asymptotic properties of fields and space-times}},
  \href{http://dx.doi.org/10.1103/PhysRevLett.10.66}{\emph{Phys. Rev. Lett.}
  {\bf 10} (1963) 66--68}.

\bibitem{Penrose:1964wq}
R.~Penrose, \emph{{Gravitational collapse and space-time singularities}},
  \href{http://dx.doi.org/10.1103/PhysRevLett.14.57}{\emph{Phys. Rev. Lett.}
  {\bf 14} (1965) 57--59}.

\bibitem{Penrose:1969pc}
R.~Penrose, \emph{{Gravitational collapse: The role of general relativity}},
  {\emph{Riv. Nuovo Cim.} {\bf 1} (1969) 252--276}.

\bibitem{Wald:1997wa}
R.~M. Wald, \emph{{Gravitational collapse and cosmic censorship}},  pp.~69--85,
  10, 1997.
\newblock \href{http://arxiv.org/abs/gr-qc/9710068}{{\tt gr-qc/9710068}}.
\newblock \href{http://dx.doi.org/10.1007/978-94-017-0934-7_5}{DOI}.

\bibitem{Hartle:1996rp}
J.~B. Hartle, \emph{{Generalized quantum theory in evaporating black hole
  space-times}},  in \emph{{Symposium on Black Holes and Relativistic Stars
  (dedicated to memory of S. Chandrasekhar)}}, pp.~195--219, 12, 1996.
\newblock \href{http://arxiv.org/abs/gr-qc/9705022}{{\tt gr-qc/9705022}}.

\bibitem{Mathur:2005zp}
S.~D. Mathur, \emph{{The Fuzzball proposal for black holes: An Elementary
  review}}, \href{http://dx.doi.org/10.1002/prop.200410203}{\emph{Fortsch.
  Phys.} {\bf 53} (2005) 793--827},
  [\href{http://arxiv.org/abs/hep-th/0502050}{{\tt hep-th/0502050}}].

\bibitem{Cardoso:2016rao}
V.~Cardoso, E.~Franzin and P.~Pani, \emph{{Is the gravitational-wave ringdown a
  probe of the event horizon?}},
  \href{http://dx.doi.org/10.1103/PhysRevLett.116.171101}{\emph{Phys. Rev.
  Lett.} {\bf 116} (2016) 171101}, [\href{http://arxiv.org/abs/1602.07309}{{\tt
  1602.07309}}].

\bibitem{Cardoso:2017cqb}
V.~Cardoso and P.~Pani, \emph{{Tests for the existence of black holes through
  gravitational wave echoes}},
  \href{http://dx.doi.org/10.1038/s41550-017-0225-y}{\emph{Nature Astron.} {\bf
  1} (2017) 586--591}, [\href{http://arxiv.org/abs/1709.01525}{{\tt
  1709.01525}}].

\bibitem{Bianchi:2018kzy}
M.~Bianchi, D.~Consoli, A.~Grillo and J.~F. Morales, \emph{{The dark side of
  fuzzball geometries}},
  \href{http://dx.doi.org/10.1007/JHEP05(2019)126}{\emph{JHEP} {\bf 05} (2019)
  126}, [\href{http://arxiv.org/abs/1811.02397}{{\tt 1811.02397}}].

\bibitem{Bianchi:2020des}
M.~Bianchi, A.~Grillo and J.~F. Morales, \emph{{Chaos at the rim of black hole
  and fuzzball shadows}},
  \href{http://dx.doi.org/10.1007/JHEP05(2020)078}{\emph{JHEP} {\bf 05} (2020)
  078}, [\href{http://arxiv.org/abs/2002.05574}{{\tt 2002.05574}}].

\bibitem{Bianchi:2020yzr}
M.~Bianchi, D.~Consoli, A.~Grillo and J.~F. Morales, \emph{{Light rings of
  five-dimensional geometries}},
  \href{http://dx.doi.org/10.1007/JHEP03(2021)210}{\emph{JHEP} {\bf 03} (2021)
  210}, [\href{http://arxiv.org/abs/2011.04344}{{\tt 2011.04344}}].

\bibitem{Bena:2020see}
I.~Bena and D.~R. Mayerson, \emph{{Multipole Ratios: A New Window into Black
  Holes}}, \href{http://dx.doi.org/10.1103/PhysRevLett.125.221602}{\emph{Phys.
  Rev. Lett.} {\bf 125} (2020) 22},
  [\href{http://arxiv.org/abs/2006.10750}{{\tt 2006.10750}}].

\bibitem{Bianchi:2020bxa}
M.~Bianchi, D.~Consoli, A.~Grillo, J.~F. Morales, P.~Pani and G.~Raposo,
  \emph{{Distinguishing fuzzballs from black holes through their multipolar
  structure}},
  \href{http://dx.doi.org/10.1103/PhysRevLett.125.221601}{\emph{Phys. Rev.
  Lett.} {\bf 125} (2020) 221601}, [\href{http://arxiv.org/abs/2007.01743}{{\tt
  2007.01743}}].

\bibitem{Bena:2020uup}
I.~Bena and D.~R. Mayerson, \emph{{Black Holes Lessons from Multipole Ratios}},
  \href{http://dx.doi.org/10.1007/JHEP03(2021)114}{\emph{JHEP} {\bf 03} (2021)
  114}, [\href{http://arxiv.org/abs/2007.09152}{{\tt 2007.09152}}].

\bibitem{Bianchi:2020miz}
M.~Bianchi, D.~Consoli, A.~Grillo, J.~F. Morales, P.~Pani and G.~Raposo,
  \emph{{The multipolar structure of fuzzballs}},
  \href{http://dx.doi.org/10.1007/JHEP01(2021)003}{\emph{JHEP} {\bf 01} (2021)
  003}, [\href{http://arxiv.org/abs/2008.01445}{{\tt 2008.01445}}].

\bibitem{Mayerson:2020tpn}
D.~R. Mayerson, \emph{{Fuzzballs and Observations}},
  \href{http://dx.doi.org/10.1007/s10714-020-02769-w}{\emph{Gen. Rel. Grav.}
  {\bf 52} (2020) 115}, [\href{http://arxiv.org/abs/2010.09736}{{\tt
  2010.09736}}].

\bibitem{Bah:2021jno}
I.~Bah, I.~Bena, P.~Heidmann, Y.~Li and D.~R. Mayerson, \emph{{Gravitational
  Footprints of Black Holes and Their Microstate Geometries}},
  \href{http://arxiv.org/abs/2104.10686}{{\tt 2104.10686}}.

\bibitem{Ikeda:2021uvc}
T.~Ikeda, M.~Bianchi, D.~Consoli, A.~Grillo, J.~F. Morales, P.~Pani et~al.,
  \emph{{Black-hole microstate spectroscopy: ringdown, quasinormal modes, and
  echoes}},  \href{http://arxiv.org/abs/2103.10960}{{\tt 2103.10960}}.

\bibitem{Cardoso:2019rvt}
V.~Cardoso and P.~Pani, \emph{{Testing the nature of dark compact objects: a
  status report}},
  \href{http://dx.doi.org/10.1007/s41114-019-0020-4}{\emph{Living Rev. Rel.}
  {\bf 22} (2019) 4}, [\href{http://arxiv.org/abs/1904.05363}{{\tt
  1904.05363}}].

\bibitem{Love}
A.~E.~H. Love, \emph{{The yielding of the earth to disturbing forces}},
  \href{http://dx.doi.org/10.1098/rspa.1909.0008.}{\emph{Proceedings of the
  Royal Society of London Series A} {\bf 82} (1909) 73}.

\bibitem{Flanagan:2007ix}
E.~E. Flanagan and T.~Hinderer, \emph{{Constraining neutron star tidal Love
  numbers with gravitational wave detectors}},
  \href{http://dx.doi.org/10.1103/PhysRevD.77.021502}{\emph{Phys. Rev. D} {\bf
  77} (2008) 021502}, [\href{http://arxiv.org/abs/0709.1915}{{\tt 0709.1915}}].

\bibitem{Damour:2009vw}
T.~Damour and A.~Nagar, \emph{{Relativistic tidal properties of neutron
  stars}}, \href{http://dx.doi.org/10.1103/PhysRevD.80.084035}{\emph{Phys. Rev.
  D} {\bf 80} (2009) 084035}, [\href{http://arxiv.org/abs/0906.0096}{{\tt
  0906.0096}}].

\bibitem{Binnington:2009bb}
T.~Binnington and E.~Poisson, \emph{{Relativistic theory of tidal Love
  numbers}}, \href{http://dx.doi.org/10.1103/PhysRevD.80.084018}{\emph{Phys.
  Rev. D} {\bf 80} (2009) 084018}, [\href{http://arxiv.org/abs/0906.1366}{{\tt
  0906.1366}}].

\bibitem{Fang:2005qq}
H.~Fang and G.~Lovelace, \emph{{Tidal coupling of a Schwarzschild black hole
  and circularly orbiting moon}},
  \href{http://dx.doi.org/10.1103/PhysRevD.72.124016}{\emph{Phys. Rev. D} {\bf
  72} (2005) 124016}, [\href{http://arxiv.org/abs/gr-qc/0505156}{{\tt
  gr-qc/0505156}}].

\bibitem{Kol:2011vg}
B.~Kol and M.~Smolkin, \emph{{Black hole stereotyping: Induced gravito-static
  polarization}}, \href{http://dx.doi.org/10.1007/JHEP02(2012)010}{\emph{JHEP}
  {\bf 02} (2012) 010}, [\href{http://arxiv.org/abs/1110.3764}{{\tt
  1110.3764}}].

\bibitem{Hui:2020xxx}
L.~Hui, A.~Joyce, R.~Penco, L.~Santoni and A.~R. Solomon, \emph{{Static
  response and Love numbers of Schwarzschild black holes}},
  \href{http://dx.doi.org/10.1088/1475-7516/2021/04/052}{\emph{JCAP} {\bf 04}
  (2021) 052}, [\href{http://arxiv.org/abs/2010.00593}{{\tt 2010.00593}}].

\bibitem{Pereniguez:2021xcj}
D.~Pere\~niguez and V.~Cardoso, \emph{{Love numbers and magnetic susceptibility
  of charged black holes}},
  \href{http://dx.doi.org/10.1103/PhysRevD.105.044026}{\emph{Phys. Rev. D} {\bf
  105} (2022) 044026}, [\href{http://arxiv.org/abs/2112.08400}{{\tt
  2112.08400}}].

\bibitem{Cardoso:2017cfl}
V.~Cardoso, E.~Franzin, A.~Maselli, P.~Pani and G.~Raposo, \emph{{Testing
  strong-field gravity with tidal Love numbers}},
  \href{http://dx.doi.org/10.1103/PhysRevD.95.084014}{\emph{Phys. Rev. D} {\bf
  95} (2017) 084014}, [\href{http://arxiv.org/abs/1701.01116}{{\tt
  1701.01116}}].

\bibitem{Emparan:2017qxd}
R.~Emparan, A.~Fernandez-Pique and R.~Luna, \emph{{Geometric polarization of
  plasmas and Love numbers of AdS black branes}},
  \href{http://dx.doi.org/10.1007/JHEP09(2017)150}{\emph{JHEP} {\bf 09} (2017)
  150}, [\href{http://arxiv.org/abs/1707.02777}{{\tt 1707.02777}}].

\bibitem{Cardoso:2018ptl}
V.~Cardoso, M.~Kimura, A.~Maselli and L.~Senatore, \emph{{Black Holes in an
  Effective Field Theory Extension of General Relativity}},
  \href{http://dx.doi.org/10.1103/PhysRevLett.121.251105}{\emph{Phys. Rev.
  Lett.} {\bf 121} (2018) 251105}, [\href{http://arxiv.org/abs/1808.08962}{{\tt
  1808.08962}}].

\bibitem{Chakraborty:2023zed}
S.~Chakraborty, E.~Maggio, M.~Silvestrini and P.~Pani, \emph{{Dynamical tidal
  Love numbers of Kerr-like compact objects}},
  \href{http://arxiv.org/abs/2310.06023}{{\tt 2310.06023}}.

\bibitem{Piovano:2022ojl}
G.~A. Piovano, A.~Maselli and P.~Pani, \emph{{Constraining the tidal
  deformability of supermassive objects with extreme mass ratio inspirals and
  semianalytical frequency-domain waveforms}},
  \href{http://dx.doi.org/10.1103/PhysRevD.107.024021}{\emph{Phys. Rev. D} {\bf
  107} (2023) 024021}, [\href{http://arxiv.org/abs/2207.07452}{{\tt
  2207.07452}}].

\bibitem{Fucito:2023afe}
F.~Fucito and J.~F. Morales, \emph{{Post Newtonian emission of gravitational
  waves from binary systems: a gauge theory perspective}},
  \href{http://arxiv.org/abs/2311.14637}{{\tt 2311.14637}}.

\bibitem{Aminov:2020yma}
G.~Aminov, A.~Grassi and Y.~Hatsuda, \emph{{Black Hole Quasinormal Modes and
  Seiberg-Witten Theory}},  \href{http://arxiv.org/abs/2006.06111}{{\tt
  2006.06111}}.

\bibitem{Bonelli:2021uvf}
G.~Bonelli, C.~Iossa, D.~P. Lichtig and A.~Tanzini, \emph{{Exact solution of
  Kerr black hole perturbations via CFT2 and instanton counting: Greybody
  factor, quasinormal modes, and Love numbers}},
  \href{http://dx.doi.org/10.1103/PhysRevD.105.044047}{\emph{Phys. Rev. D} {\bf
  105} (2022) 044047}, [\href{http://arxiv.org/abs/2105.04483}{{\tt
  2105.04483}}].

\bibitem{Bianchi:2021xpr}
M.~Bianchi, D.~Consoli, A.~Grillo and J.~F. Morales, \emph{{QNMs of branes, BHs
  and fuzzballs from quantum SW geometries}},
  \href{http://dx.doi.org/10.1016/j.physletb.2021.136837}{\emph{Phys. Lett. B}
  {\bf 824} (2022) 136837}, [\href{http://arxiv.org/abs/2105.04245}{{\tt
  2105.04245}}].

\bibitem{Bianchi:2021mft}
M.~Bianchi, D.~Consoli, A.~Grillo and J.~F. Morales, \emph{{More on the SW-QNM
  correspondence}},
  \href{http://dx.doi.org/10.1007/JHEP01(2022)024}{\emph{JHEP} {\bf 01} (2022)
  024}, [\href{http://arxiv.org/abs/2109.09804}{{\tt 2109.09804}}].

\bibitem{Bonelli:2022ten}
G.~Bonelli, C.~Iossa, D.~P. Lichtig and A.~Tanzini, \emph{{Irregular Liouville
  correlators and connection formulae for Heun functions}},
  \href{http://arxiv.org/abs/2201.04491}{{\tt 2201.04491}}.

\bibitem{Consoli:2022eey}
D.~Consoli, F.~Fucito, J.~F. Morales and R.~Poghossian, \emph{{CFT description
  of BH\textquoteright{}s and ECO\textquoteright{}s: QNMs, superradiance,
  echoes and tidal responses}},
  \href{http://dx.doi.org/10.1007/JHEP12(2022)115}{\emph{JHEP} {\bf 12} (2022)
  115}, [\href{http://arxiv.org/abs/2206.09437}{{\tt 2206.09437}}].

\bibitem{Bianchi:2021yqs}
M.~Bianchi and G.~Di~Russo, \emph{{Turning black-holes and D-branes inside out
  their photon-spheres}},  \href{http://arxiv.org/abs/2110.09579}{{\tt
  2110.09579}}.

\bibitem{Bianchi:2022wku}
M.~Bianchi and G.~Di~Russo, \emph{{Turning rotating D-branes and BHs inside out
  their photon-halo}},  \href{http://arxiv.org/abs/2203.14900}{{\tt
  2203.14900}}.

\bibitem{Bianchi:2022qph}
M.~Bianchi and G.~Di~Russo, \emph{{2-charge circular fuzz-balls and their
  perturbations}}, \href{http://dx.doi.org/10.1007/JHEP08(2023)217}{\emph{JHEP}
  {\bf 08} (2023) 217}, [\href{http://arxiv.org/abs/2212.07504}{{\tt
  2212.07504}}].

\bibitem{Aminov:2023jve}
G.~Aminov, P.~Arnaudo, G.~Bonelli, A.~Grassi and A.~Tanzini, \emph{{Black hole
  perturbation theory and multiple polylogarithms}},
  \href{http://arxiv.org/abs/2307.10141}{{\tt 2307.10141}}.

\bibitem{Bianchi:2023sfs}
M.~Bianchi, G.~Di~Russo, A.~Grillo, J.~F. Morales and G.~Sudano, \emph{{On the
  stability and deformability of top stars}},
  \href{http://arxiv.org/abs/2305.15105}{{\tt 2305.15105}}.

\bibitem{Bautista:2023sdf}
Y.~F. Bautista, G.~Bonelli, C.~Iossa, A.~Tanzini and Z.~Zhou, \emph{{Black Hole
  Perturbation Theory Meets CFT$_2$: Kerr Compton Amplitudes from
  Nekrasov-Shatashvili Functions}},
  \href{http://arxiv.org/abs/2312.05965}{{\tt 2312.05965}}.

\bibitem{Seiberg:1994rs}
N.~Seiberg and E.~Witten, \emph{{Electric - magnetic duality, monopole
  condensation, and confinement in N=2 supersymmetric Yang-Mills theory}},
  \href{http://dx.doi.org/10.1016/0550-3213(94)90124-4}{\emph{Nucl. Phys. B}
  {\bf 426} (1994) 19--52}, [\href{http://arxiv.org/abs/hep-th/9407087}{{\tt
  hep-th/9407087}}].

\bibitem{Nekrasov:2002qd}
N.~A. Nekrasov, \emph{{Seiberg-Witten prepotential from instanton counting}},
  \href{http://dx.doi.org/10.4310/ATMP.2003.v7.n5.a4}{\emph{Adv. Theor. Math.
  Phys.} {\bf 7} (2003) 831--864},
  [\href{http://arxiv.org/abs/hep-th/0206161}{{\tt hep-th/0206161}}].

\bibitem{Flume:2002az}
R.~Flume and R.~Poghossian, \emph{{An Algorithm for the microscopic evaluation
  of the coefficients of the Seiberg-Witten prepotential}},
  \href{http://dx.doi.org/10.1142/S0217751X03013685}{\emph{Int. J. Mod. Phys.
  A} {\bf 18} (2003) 2541}, [\href{http://arxiv.org/abs/hep-th/0208176}{{\tt
  hep-th/0208176}}].

\bibitem{Bruzzo:2002xf}
U.~Bruzzo, F.~Fucito, J.~F. Morales and A.~Tanzini, \emph{{Multiinstanton
  calculus and equivariant cohomology}},
  \href{http://dx.doi.org/10.1088/1126-6708/2003/05/054}{\emph{JHEP} {\bf 05}
  (2003) 054}, [\href{http://arxiv.org/abs/hep-th/0211108}{{\tt
  hep-th/0211108}}].

\bibitem{Alday:2009aq}
L.~F. Alday, D.~Gaiotto and Y.~Tachikawa, \emph{{Liouville Correlation
  Functions from Four-dimensional Gauge Theories}},
  \href{http://dx.doi.org/10.1007/s11005-010-0369-5}{\emph{Lett. Math. Phys.}
  {\bf 91} (2010) 167--197}, [\href{http://arxiv.org/abs/0906.3219}{{\tt
  0906.3219}}].

\bibitem{Lunin:2001fv}
O.~Lunin and S.~D. Mathur, \emph{{Metric of the multiply wound rotating
  string}}, \href{http://dx.doi.org/10.1016/S0550-3213(01)00321-2}{\emph{Nucl.
  Phys. B} {\bf 610} (2001) 49--76},
  [\href{http://arxiv.org/abs/hep-th/0105136}{{\tt hep-th/0105136}}].

\bibitem{Bah:2020ogh}
I.~Bah and P.~Heidmann, \emph{{Topological Stars and Black Holes}},
  \href{http://dx.doi.org/10.1103/PhysRevLett.126.151101}{\emph{Phys. Rev.
  Lett.} {\bf 126} (2021) 151101}, [\href{http://arxiv.org/abs/2011.08851}{{\tt
  2011.08851}}].

\bibitem{Bah:2020pdz}
I.~Bah and P.~Heidmann, \emph{{Topological stars, black holes and generalized
  charged Weyl solutions}},
  \href{http://dx.doi.org/10.1007/JHEP09(2021)147}{\emph{JHEP} {\bf 09} (2021)
  147}, [\href{http://arxiv.org/abs/2012.13407}{{\tt 2012.13407}}].

\bibitem{Bena:2015bea}
I.~Bena, S.~Giusto, R.~Russo, M.~Shigemori and N.~P. Warner, \emph{{Habemus
  Superstratum! A constructive proof of the existence of superstrata}},
  \href{http://dx.doi.org/10.1007/JHEP05(2015)110}{\emph{JHEP} {\bf 05} (2015)
  110}, [\href{http://arxiv.org/abs/1503.01463}{{\tt 1503.01463}}].

\bibitem{Bena:2017xbt}
I.~Bena, S.~Giusto, E.~J. Martinec, R.~Russo, M.~Shigemori, D.~Turton et~al.,
  \emph{{Asymptotically-flat supergravity solutions deep inside the black-hole
  regime}}, \href{http://dx.doi.org/10.1007/JHEP02(2018)014}{\emph{JHEP} {\bf
  02} (2018) 014}, [\href{http://arxiv.org/abs/1711.10474}{{\tt 1711.10474}}].

\bibitem{Nekrasov:2009rc}
N.~A. Nekrasov and S.~L. Shatashvili, \emph{{Quantization of Integrable Systems
  and Four Dimensional Gauge Theories}},  in \emph{{16th International Congress
  on Mathematical Physics}}, pp.~265--289, 8, 2009.
\newblock \href{http://arxiv.org/abs/0908.4052}{{\tt 0908.4052}}.
\newblock \href{http://dx.doi.org/10.1142/9789814304634_0015}{DOI}.

\bibitem{Poghossian:2010pn}
R.~Poghossian, \emph{{Deforming SW curve}},
  \href{http://dx.doi.org/10.1007/JHEP04(2011)033}{\emph{JHEP} {\bf 04} (2011)
  033}, [\href{http://arxiv.org/abs/1006.4822}{{\tt 1006.4822}}].

\bibitem{Fucito:2011pn}
F.~Fucito, J.~F. Morales, D.~R. Pacifici and R.~Poghossian, \emph{{Gauge
  theories on $\Omega$-backgrounds from non commutative Seiberg-Witten
  curves}}, \href{http://dx.doi.org/10.1007/JHEP05(2011)098}{\emph{JHEP} {\bf
  05} (2011) 098}, [\href{http://arxiv.org/abs/1103.4495}{{\tt 1103.4495}}].

\bibitem{Poghosyan:2020zzg}
H.~Poghosyan, \emph{{Recursion relation for instanton counting for SU(2) $
  \mathcal{N} $ = 2 SYM in NS limit of $\Omega$ background}},
  \href{http://dx.doi.org/10.1007/JHEP05(2021)088}{\emph{JHEP} {\bf 05} (2021)
  088}, [\href{http://arxiv.org/abs/2010.08498}{{\tt 2010.08498}}].

\bibitem{Bena:2017upb}
I.~Bena, D.~Turton, R.~Walker and N.~P. Warner, \emph{{Integrability and
  Black-Hole Microstate Geometries}},
  \href{http://dx.doi.org/10.1007/JHEP11(2017)021}{\emph{JHEP} {\bf 11} (2017)
  021}, [\href{http://arxiv.org/abs/1709.01107}{{\tt 1709.01107}}].

\bibitem{Giusto:2023awo}
S.~Giusto, C.~Iossa and R.~Russo, \emph{{The black hole behind the cut}},
  \href{http://dx.doi.org/10.1007/JHEP10(2023)050}{\emph{JHEP} {\bf 10} (2023)
  050}, [\href{http://arxiv.org/abs/2306.15305}{{\tt 2306.15305}}].

\end{thebibliography}
\end{document}